\documentstyle[prl,aps,multicol,epsfig,amssymb]{revtex}

\begin{document}

\title{Accurate effective pair potentials for polymer solutions}
\author {P.G. Bolhuis, A.A. Louis, and J.P. Hansen}
 \address{Department of Chemistry, Lensfield Rd,
Cambridge CB2 1EW, UK}
\author{E.J. Meijer }
 \address{Department of Chemical
Engineering, University of Amsterdam, Nieuwe Achtergracht 166, NL-1018
WV Amsterdam, Netherlands.}  \date{\today} \maketitle
\begin{abstract}
\noindent 
Dilute or semi-dilute solutions of non-intersecting self-avoiding walk
(SAW) polymer chains are mapped onto a fluid of ``soft'' particles
interacting via an effective pair potential between their centers of
mass. This mapping is achieved by inverting the pair distribution
function of the centers of mass of the original polymer chains, using
integral equation techniques from the theory of simple fluids. The
resulting effective pair potential is finite at all distances, has a
range of the order of the radius of gyration, and turns out to be only
moderately concentration-dependent.  The dependence of the effective
potential on polymer length is analyzed in an effort to extract the
scaling limit. The effective potential is used to derive the osmotic
equation of state, which is compared to simulation data for the full
SAW segment model, and to the predictions of renormalization group
calculations. A similar inversion procedure is used to derive an
effective wall-polymer potential from the center of mass density
profiles near the wall, obtained from simulations of the full polymer
segment model. The resulting wall-polymer potential turns out to
depend strongly on bulk polymer concentration when polymer-polymer
correlations are taken into account, leading to a considerable
enhancement of the effective repulsion with increasing
concentration. The effective polymer-polymer and wall-polymer
potentials are combined to calculate the depletion interaction induced
by SAW polymers between two walls. The calculated depletion
interaction agrees well with the ``exact'' results from much more
computer-intensive direct simulation of the full polymer-segment
model, and clearly illustrates the inadequacy -- in the semi-dilute
regime -- of the standard Asakura-Oosawa approximation based on the
assumption of non-interacting polymer coils.
\end{abstract}
%\vspace*{-0.2cm}
\pacs{61.25.H,61.20.Gy,82.70Dd}
%61.25.H Polymer solutions
%61.20.-p Structure of liquids
%82.70Dd Colloids
\begin{multicols}{2}

\section{Introduction}

\begin{figure}[tbp]
\begin{center}
\epsfig{figure=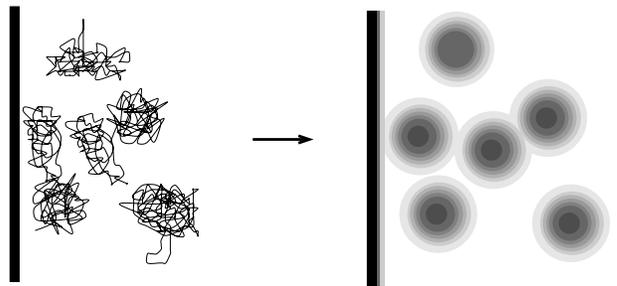,width=8cm}
\caption{\label{fig:effective} Modelling polymer coils by effective
``soft particles''.  The $N$ polymers, each made up of $L$ segments, are
replaced by $N$ particles interacting with an effective pair
potential.  The centers of the particles correspond to the polymer
CM.  The interaction of the polymers with a hard wall
is modelled by a single soft particle--wall interaction.}
\end{center}
\end{figure}

Polymer solutions have attracted the attention of theorists and
experimentalists alike for many decades, and a theoretical
understanding of their structure and phase behavior, based on
mean-field and on scaling or renormalization group arguments, is by
now well
established~\cite{Flor53,deGe79,Doi86,desC90,Gros94,Doi95}. Recently,
there has been a growing interest in the structure, phase behavior and
rheology of binary systems involving colloidal particles and
non-adsorbing
polymer~\cite{Lekk92,Meij94,Dick94,Ilet95,Ohsh97,Verm98,Mous99,Poon99,Bech99,Weiss99,Kulk99,Hank99,Loui99a,deHo99,Brad00,Tuin00a}.
In such mixtures the mean size of the polymer coils, i.e.\ their radius
of gyration $R_g$, is comparable to, or smaller than the diameter
$\sigma$ of the colloidal particles. Since the latter may, for most
purposes, be modeled as ``hard'' convex bodies dominated by excluded
volume effects on the mesoscopic scale $\sigma$, it is clear that a
statistical description of the polymer coils requires a high degree of
coarse-graining to provide a tractable theory of these mixtures. Such
coarse-graining is, more generally, desirable for theoretical
investigations of large scale phenomena involving large numbers of
interacting polymer chains  in the dilute of semi-dilute regimes. In
particular, simulations of solutions involving many interacting
polymer chains become rapidly intractable if a detailed description at
the level of monomers or even of Kuhn segments is retained. It is
therefore tempting to consider polymer coils as ``soft'' particles,
and to replace the detailed interactions between segments by an
effective interaction acting between the centers of mass (CM) of different
polymer coils as shown schematically in
Fig.~\ref{fig:effective}.
 Similarly, an effective interaction must be
worked out between the ``soft'' polymer coils and the ``hard''
colloidal particles.  Such a drastic reduction in the number of
degrees of freedom, achieved by formally averaging over the
coordinates of individual polymer segments, leads to a considerable
simplification of the initial problem involving $N_c$ colloidal
particles and $N_p L$ polymer segments, where $N_p$ is the number of
polymer coils and $L$ the number of monomers or segments per polymer
(i.e.\ the length of a polymer).  The idea of representing a polymer
coil by a single particle of radius of the order of $R_g$ goes back to
the work of Flory and Krigbaum~\cite{Flor50}, who considered the
infinite dilution limit of two isolated interacting polymers. A brief
outline of subsequent theoretical and numerical work on the two-coil
problem is given in Section~\ref{sec:twopol}. In this paper we
generalize the idea to finite concentrations, i.e.\ to dilute and
semi-dilute polymer solutions. The effective interaction between the
CM of polymer coils is determined by a combination of
Monte Carlo (MC) simulations of a detailed segment model of interacting
polymers, and of an inversion technique which allows the effective
pair interaction to be extracted from the MC results for the center of
mass pair distribution function. A similar inversion technique is
applied to the density profiles of the CM  of the polymers
near a hard wall to determine the effective interaction between a
wall, impenetrable to the polymer segments, and the CM  of
interacting polymers. The effective polymer-polymer and wall-polymer
interactions provide a first step towards a complete description of
colloid-polymer mixtures, with the hard wall considered in this paper
representing a single colloidal particle of infinite radius. The
ultimate goal is to go well beyond the familiar Asakura-Oosawa (AO)
model which considers polymers to be non-interacting point particles,
excluded from a sphere of radius $\sigma/2 + R_g$ around each
colloidal particle~\cite{Asak58}. This model leads to the well-known
AO depletion interaction between hard sphere
colloids~\cite{Asak58,Asak54,LiIn75}. As an application of the general
method outlined in this paper, the limitations of the AO picture will
be illustrated in a calculation of the depletion interaction between
two parallel hard walls. The effective interaction between polymer
coils will be shown to lead to considerable deviations from the AO
results, even in the dilute regime.

A preliminary account of parts of the present work has been published
elsewhere~\cite{Loui00}. A related soft particle picture has
recently been applied to polymer melts and polymer
blends~\cite{Mura98}. However, the phenomenological coarse-graining
procedure proposed by these authors, and its practical implementation,
differ considerably from the present ``first principles'' approach,
which is better adapted to dilute and semi-dilute polymer
solutions. Both methods are good examples of current efforts to bridge
widely different length and time scales in complex fluids.

\section{Simulation Models and Methods}
\label{sec:model}

Many physical properties of polymers in solution already emerge from
simple models which ignore chemical detail and describe the polymers
as self avoiding walks (SAW) with hard segments interacting through a
simple potential.  For example, solutions of linear polymers in a good
solvent are well modeled by $N$ athermal SAW's, each made up of $L$
non-intersecting segments, on a cubic lattice of $M$ sites, with
periodic boundary conditions.  This model captures the leading scaling
behavior and has been used for many decades to describe polymer
solutions\cite{Flor53,deGe79,Doi86,desC90,Gros94,Doi95}.  Slightly
more sophisticated models exist, such as the fluctuating bond
model~\cite{flucbond} or off-lattice hard sphere chains\cite{Daut94},
but the SAW lattice model is simple, efficient and allows for
comparisons with previous studies.

Within the lattice model, the monomer packing fraction is equal to the
fraction of lattice sites occupied by polymer segments, $c=N\times
L/M$, while the concentration of polymer chains is $\rho_b=c/L=N/M$.
For a single SAW chain, the radius of gyration scales as $R_g \sim
L^{\nu}$, where $\nu \simeq 0.6$ is the Flory exponent\cite{Flor53}.
The overlap concentration $\rho^*$, signalling the onset of the
semi-dilute regime, is such that $4 \pi \rho^*R_g^3/3 \simeq 1$, and
hence $\rho^* \sim L^{-3 \nu}$\cite{semidilute}.

To sample the configuration space of the polymer system we employ the
Monte Carlo pivot algorithm\cite{Daut94,pivot} which attempts to
rotate part of the polymer around a random segment (the pivot). If the
new trial configuration shows no overlap, the move is accepted,
otherwise the old configuration is restored.  This simple scheme turns
out to be very effective for single polymers and dilute polymer
solutions where we found that it efficiently samples configurational
space up to densities $\rho_b/\rho^*\approx 1$ for $L=500$ polymers.
Because the polymers are restricted to a cubic lattice, the pivot move
can only take place in 5 possible directions.  For efficiency we store
the complete lattice in memory, so that overlap between different
polymers can be easily checked for. In this way one has only to check
of order $L$ sites per polymer move, which is much more efficient than
the $N L^2$ sites needed when each pair of segments has to be tested
for overlap.

In addition to the pivot moves, we also attempt to translate the
polymer.  This Monte Carlo  move enhances the relaxation to
equilibrium of the polymer solution, although the acceptance ratio for
this move decreases rapidly if the density exceeds $\rho_b/\rho^*
\approx1$ (for $L=500$ polymers).  For densities deep in the
semi-dilute regime, $\rho_b/\rho^*>1$, we therefore also perform
configurational bias Monte Carlo (CBMC)
moves~\cite{frenkelbook,dijkstra}, in which part of the interior
polymer is regrown. In addition, we attempt reptation moves where a
limited number of segments at one end of the polymer are removed and
regrown at the other end. By regrowing the polymer a bias is
introduced, which is then corrected for in the
sampling~\cite{frenkelbook,dijkstra}.  In the simulations at high
densities, we find that we can regrow groups of  up to about 20-40
segments in a CBMC move with a reasonable acceptance ratio (about 40 -
50~\%). More sophisticated algorithms for very dense polymer systems
are available~\cite{vlugt}, but are not necessary in our relatively
dilute systems.

\section{Effective potentials: two isolated polymers}
\label{sec:twopol}

The theory of the effective interaction between two polymer coils in
dilute solution has a long history.  The first calculations were by
Flory and Krigbaum in 1950\cite{Flor50}, who showed that, within a
mean-field picture, SAW polymers in a good solvent have a strongly
repulsive interaction of the form: 
\begin{equation}\label{eqII.1}
 \beta v_2^{(FK)}(r) \sim L^2\left(\frac{3}{4 \pi R_g^3}\right) (1-2
\chi) \exp \left(-\frac{3}{4} \frac{r^2}{R_g^2}\right),
\end{equation}
where $r$ is the distance between the CM  of the two
polymer coils, $\chi$ is the usual Flory parameter, and $ \beta =
1/k_{\rm B}T$ is the reciprocal temperature, with $k_{\rm B}$ denoting
Boltzmann's constant.  As long as the polymers are in a good solvent,
the chains can be regarded as athermal, and for that reason we set $\beta
= 1$ in the rest of this paper.

The interaction strength at full overlap,
$  v_2^{(FK)}(r=0)$, can be understood from the following
argument: each polymer coil has a density of monomers $c \sim L/V$,
while the volume of a polymer scales as $V \propto L^{3\nu}$, so that
$c \propto L^{1 - 3 \nu}$ (here $\nu \approx 0.6$ is the Flory
exponent).  If two polymers overlap completely then the mean-field
free-energy of interaction would be proportional to the number of
monomers times the probability of contact of two monomers on different
chains:
\begin{equation}\label{eq1.1}
  v_2^{(FK)}(r=0) \propto L c \propto L^{2-3 \nu} \sim {\cal O}(L^{0.2}),
\end{equation}
which implies that the polymer repulsion increases with polymerization
$L$, and is typically much larger than $k_BT$.

In an elegant paper, Grosberg, Khalatur, and Khokhlov\cite{Gros82}
showed that Flory's argument was in fact incorrect.  From scaling
theory it follows that the probability of an interaction between two
monomers on different chains scales as $c^{1/(3 \nu-1)} \sim
c^{1.3}$\cite{Daou75} instead of simply $c$, so that the free energy
of interaction scales as: 
\begin{equation}\label{eq1.2}
   v_2(r=0) \propto L c^{1/(3 \nu -1)} \sim L (L^{1-3\nu})^{1/(3
\nu -1)} \sim {\cal O}(1).
\end{equation}
In other words, the free energy of interaction at full overlap of two
equal-length polymers is independent of the degree of polymerization;
polymer coils are not nearly as ``hard'' as one might naively expect.

\begin{figure}
\begin{center}
\epsfig{figure=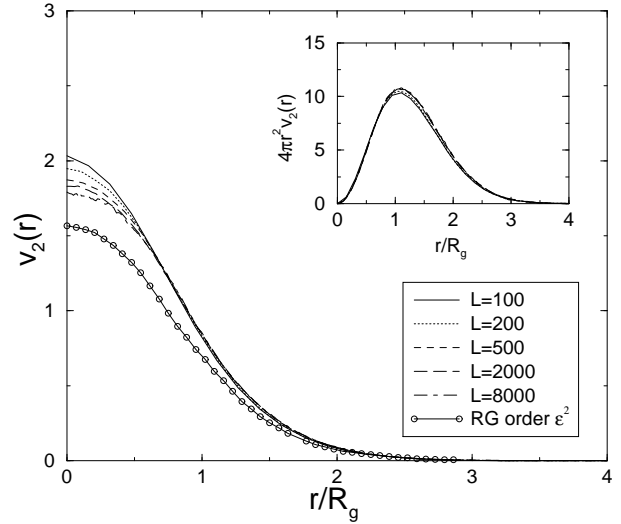,width=8cm}
\caption{\label{fig:v0}
Effective CM-CM pair potential $v_2(r)$ for two isolated SAW polymers,
here shown for different lengths $L$. The
x-axis is scaled with $R_g$, to allow comparison. The pair interaction
$v_2(r)$ is approximately Gaussian. The height of the potential at
$r=0$ decreases with length.  Also shown is the RG result from an order
${\cal O}(\epsilon^2)$ expansion.  Inset: $4 \pi r^2 v_2(r)$, which is
more relevant to the thermodynamics of polymer solutions, shows much
less variation with length $L$ than $v_2(r)$.}
\end{center}
\end{figure}

Kr\"{u}ger, Sch\"{a}fer, and Baumg\"{a}rtner\cite{Krug89} put these
ideas on a firmer footing using elaborate renormalization group (RG)
calculations.  In particular they calculated the full free-energy of
overlap
%$  F =\exp(  v_2(r))$ 
of polymers as a function of the CM distance.  They found $ v_2(r=0) =
1.53 \epsilon$ from an r-space $\epsilon$ expansion and $ v_2(r=0) =
0.94 \epsilon + 0.62 \epsilon^2$ from a $k$-space $\epsilon$ expansion
($\epsilon = 4-d$, so $\epsilon = 1$ for 3 dimensions).  Although
there are still significant quantitative differences between an ${\cal
O}(\epsilon)$ and an ${\cal O}(\epsilon^2)$ calculation, implying that
the $\epsilon$ expansion has not quite converged, the qualitative
picture is clearly that of a repulsive Gaussian type potential, as
shown in Fig.~\ref{fig:v0}.  These calculations were confirmed by a
number of computer simulation studies, notably those by Olaj and
collaborators\cite{Olaj80}, and by Dauntenhahn and Hall\cite{Daut94}.

\begin{figure}
\begin{center}
\epsfig{figure=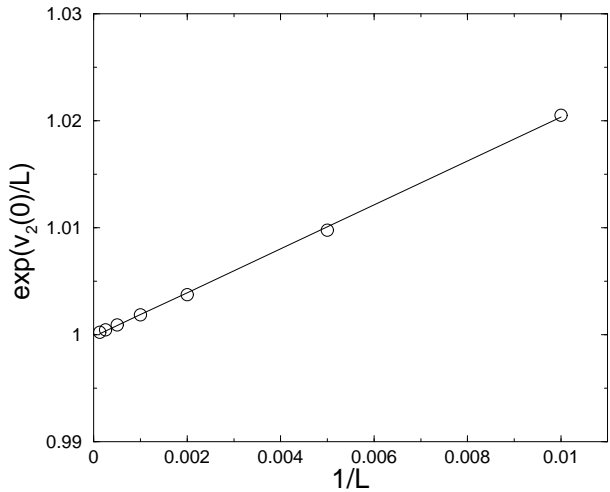,width=8cm}
\caption{\label{fig:v0fits2} Finite-size scaling for the interaction
between two isolated polymers at full overlap: $v_2(r=0)$.  By plotting
$\exp\left[v_2(r=0)/L\right]$ v.s. $1/L$, the agreement between the
scaling relation of Eq.~(\protect\ref{eq1.5}) and the simulations 
is  demonstrated.  }
\end{center}
\end{figure}

We repeated the calculation of the effective interaction between two
isolated SAW polymer coils, to make sure that the simulations are
carried out under conditions sufficiently close to the scaling limit.
For two polymers at infinite dilution, the effective interaction can be
determined by calculating the normalized probability $P(r)$ of finding
their respective CM's at a separation $r$. The effective potential
$v_2(r)$ is then defined as
\begin{equation}\label{eq3.4}
  v_2(r) = -\ln(P(r))
\end{equation}
In the course of the simulation we sample configurations of two
polymers infinitely far apart using only the pivot algorithm. After
every 1000 pivot moves, we calculate the overlap probability as a
function of CM distance, by moving the polymers towards each other
while checking for overlap.  In addition, the radius of gyration is
calculated for each length considered, from $L=100$ to $L=8000$.
This reproduces the well known Flory scaling law $R_g \sim L^{\nu}$.
The effective interactions between two polymers of various lengths are
plotted in Fig.~\ref{fig:v0}; the distance $r$ is scaled with the
measured radius of gyration $R_g$.  As expected, $ v_2(r)$ has a
Gaussian shape centered on  $r=0$.  The potentials are almost
indistinguishable for $r/R_g >1$, but for smaller $r$ the potentials
differ slightly for different $L$.  This is most pronounced at full
overlap of the polymers, where $ v_2(r=0)$ decreases with
length $L$. In the scaling limit $L\rightarrow\infty$, $ v_2(r=0)$ is
expected to reach a finite value while  for finite $L$, 
we expect $v_2(r=0)$ to scale as:
\begin{equation}
\label{eq1.5}
   v_2(r=0) \propto L \ln (1- a c^{1/(3 \nu -1)} )  \sim L \ln (1-\frac{a}{L})
\end{equation}
where $a$ is a (negative) constant and the logarithmic term arises
because $P(r)$ scales linearly.  This finite-size scaling behavior is
confirmed in Fig.\ref{fig:v0fits2} and  in the $L\rightarrow \infty$
limit this equation goes over to Eq.~(\ref{eq1.2}). Using a
non-linear fit of the MC data to Eq.~(\ref{eq1.5}) we estimate $v_2(r=0)
= 1.80 \pm 0.05$, a value slightly higher than the best ${\cal O}
(\epsilon^2)$ RG calculations which give $ v_2(r=0)=1.53$.  The
difference is most likely due to a lack of convergence of the
$\epsilon$ expansion~\cite{Krug89}.

The quantity $r^2 v_2(r)$ is actually more relevant for the
thermodynamic properties of polymer solutions than
$v_2(r=0)$\cite{Loui00a}, and, as demonstrated in the inset of
Fig.~\ref{fig:v0}, the former varies less with $L$ than the latter, such
that for $r^2 v_2(r)$ the scaling limit  appears to be reached
 even for chains as short as $L=500$.

\begin{figure}
\epsfig{figure=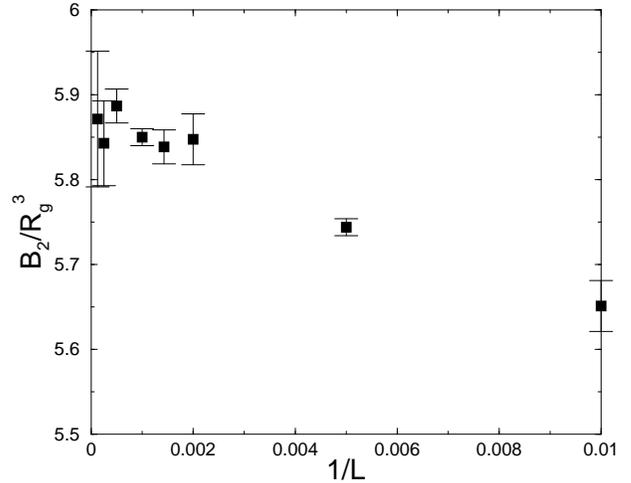,width=8cm}
\caption{\label{fig:B2} The reduced osmotic virial coefficient
$B_2/R_g^3$ v.s. $1/L$.}
\end{figure}

Similarly, with the effective pair-potentials we can calculate the 2nd
osmotic virial coefficient:
\begin{equation}\label{eq3.5}
B_2 = -2\pi \int_0^\infty r^2 dr \left( \exp(-  v_2(r)) -1
\right).
\end{equation}
Since the potentials scale with $r/R_g$, this means that $B_2/R_g^3$
should be independent of $L$ in the scaling limit.  As demonstrated in
Fig.\ref{fig:B2}, the scaling limit appears to be practically reached
for 
$L=500$.  We estimate that for $L \rightarrow \infty$, $B_2/R_g^3
\approx 5.85 \pm 0.05$, which is consistent with other results
obtained from simulations ($B_2/R_g^3 \approx 5.50$\cite{Li95}) or RG
calculations ($B_2/R_g^3 \approx 5.99$\cite{Doug84}). Note that
although $B_2$  scales as $B_2 \sim R_g^3$, as
required by scaling theory\cite{Flor53,deGe79,desC90}, this does not
imply that the polymer-polymer interaction is  hard-sphere
like, as is sometimes implied in the literature\cite{B2}.

\section{Effective Potentials: Polymer Solutions}

\subsection{Deriving effective potentials from $g(r)$}
\label{sec:polpolinv}

Having derived the effective potential between two isolated polymers,
we now attempt the same for polymers in solution at finite
concentration.  Whereas for simple fluids, the interaction potential is
generally independent of the thermodynamic state, this is  not
true for effective potentials in complex fluids.  The latter typically
follow from a coarse-graining procedure, which amounts to averaging
out certain degrees of freedom, the individual microscopic polymer
segments in the present case.  The effective total interaction
potential energy $V_N(\{{\bf r}_i\};\rho_b)$ is in fact a {\it free}
energy which depends here on the polymer density $\rho_b = N/V$ and on
the configuration $\{{\bf r}_i\}$ of the polymer CM's.  The bare
pair-interaction term $v_2(r)$ can be defined as the effective
potential between two isolated polymers, the bare triplet interaction
term $v_3({\bf r}_i,{\bf r}_j,{\bf r}_k)$ can be defined for three
isolated polymers and so forth.  One could in principle calculate
higher and higher order n-body terms, but this rapidly becomes
intractable.  Even if explicit expressions for each of the terms were
obtained, the total interaction energy would be very difficult to
evaluate because the number of n-tuple coordinates increases
exponentially.

Instead, we follow a different route and approximate the pair and
higher order terms by an effective, (state dependent) pair interaction
$ v(r;\rho_b)$ which is constructed to exactly reproduce the two-body
correlations of the full underlying many-body system.  In fact, it can
be proven that for any given pair distribution function $g(r)$ and
density $\rho_b$, there exists a corresponding {\em unique} two-body
pair potential $ v(r;\rho_b)$ which reproduces $g(r)$ {\em irrespective
of the underlying many-body interactions} in the
system\cite{Hend74}. Of course, $g(r)$ will contain contributions not
only from the bare pair-potential $v_2(r)$, but also from the three
and more body terms.  As a consequence, the effective pair
interaction $v(r;\rho_b)$ will also be state dependent (in the polymer
case, density dependent) and a new effective potential must be
calculated for each density.  Nevertheless, the effective potential
leads back to the true thermodynamics of the full many-body system
through the compressibility relation:
\begin{equation}\label{eq4.2}
\left( \frac{\partial   \Pi_b}{\partial \rho_b}\right)_{N,T} =\frac{1}{1 - \rho_b
\hat{h}(k=0)} = 1 - \rho_b \hat{c}(k=0),
\end{equation} 
where $\hat{h}(k)$ is the Fourier transform (FT) of the pair
correlation function $h(r)=g(r)-1$, and $\hat{c}(k)$ is the FT of the
direct correlation function. Using a variational argument,
Reatto\cite{Reat86} has shown that $ v(r;\rho_b)$ may also be viewed as
the ``best'' pair representation of the true interactions.  However
this inversion approach says nothing about a possible volume term
$V_0(\rho_b)$, in the coarse-grained total potential energy, which
contributes to the e.o.s.\ , but not {\em directly} to the
pair-correlations\cite{Graf98}.  Of course the volume terms may still
contribute {\em indirectly}, for example when they induce
phase-transitions.

 The inversion of $g(r)$ to extract $ v(r;\rho_b)$ is a well known
 procedure and has been studied extensively in the field of simple
 fluids\cite{Reat86,Zerah86}.  We invert $g(r)$ using the
 hypernetted-chain (HNC) closure,
\begin{equation}
\label{eq:hnc}
g(r) = exp(-  v(r) +g(r) -c(r) -1),
\end{equation}
of the Ornstein-Zernike equation\cite{Hans86}.  While the simple HNC
inversion procedure would be inadequate for dense fluids of hard core
particles, where more sophisticated closures or iterative procedures
are required\cite{Reat86,Zerah86}, we are able to demonstrate the consistency
of the HNC inversion in the present case.

\begin{figure}
\begin{center}
\epsfig{figure=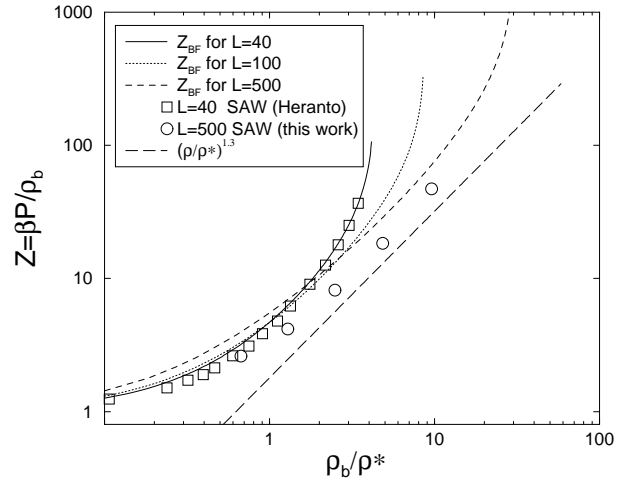,width=8cm}
\caption{\label{fig:ZBF} This figure compares the e.o.s.\ $Z= \Pi_b/\rho_b$
from simulations for $L=40$ SAW polymers\protect\cite{Hert88}, with
the e.o.s.\ for $L=500$ SAW polymers (see section~\ref{sec:eos}), and
with the Bawendi-Freed (BF) e.o.s.\ for lattice
models\protect\cite{Bawe88}, which is accurate for larger values of
$c$. The latter gives an indication where finite $c$ correlations
become important, and where one would expect the melt-regime to start
for $L=40$, $L=100$, and $L=500$ SAW polymers.  In the melt regime the
des Cloizeaux scaling law $Z \sim (\rho_b/\rho^*)^{1/(3 \nu -1)}$ will
break down\protect\cite{Doi95}. Clearly, the $L=40$ data does not follow the
des Cloizeaux scaling law, demonstrating that there is no meaningful
semi-dilute regime for $L=40$ polymers, whereas there is one for
$L=500$ polymers. }
\end{center}
\end{figure}

We performed Monte Carlo simulations of $N$ SAW polymers of length
$L=500$ on a cubic lattice of size $M = 240\times240\times240$. The
number of polymers was varied from $N=100$ ($\rho_b/\rho^*=0.54$) to
$N=6400$ ($\rho_b/\rho^* = 8.7$).  Note that at the highest density
the monomer packing fraction is $c \approx 0.23$, meaning that the
conditions for the semi-dilute regime, namely $\rho_b > \rho*$ {\em
and} $c \approx 0$ begin to be violated. At even higher densities the
system will approach the melt regime where monomer packing effects
become important\cite{prism}.  More generally, for finite length SAW
polymers, there is a limited density regime for which both conditions
for the semi-dilute regime can be simultaneously satisfied.  We find
empirically that $R_g \approx 0.39L^{0.6}$ for SAW polymers on a
simple cubic lattice, so that the monomer packing fraction at the
overlap concentration is given by
\begin{equation}
 c^* \approx 4/L^{0.8}.  
\end{equation} 
Thus, for $L=100$ chains we find $c^* \approx 0.1$ so that there is
only a very small density range which might be called semi-dilute,
while for $L=500$ chains $c^* \approx 0.027$ and a meaningful
semi-dilute regime exists. The literature contains several claims of
semi-dilute scaling behavior for SAW lattice polymers with $L<100$,
but, as the analysis above shows, these polymers do not have a
semi-dilute regime large enough to derive scaling relations.  An
example of this is shown in Fig.~\ref{fig:ZBF}.

In the course of the simulations the CM of each polymer was tracked in
order to construct the CM radial pair distribution function $g(r)$.
The latter is only known up to a cutoff radius $r_c$, which
corresponds to half the size of the simulation box (lattice size).
For the inversion, we need $g(r)$ for all $r$, so we employ the
following iterative scheme to extend $g(r)$.  First we set $g(r)=1$
for $r>r_{c}$ and calculate the corresponding $ v(r;\rho_b)$ by
inversion. We then set $ v(r;\rho_b)=0$ for $r>r_{c}$ and determine
the corresponding $g(r)$ for $0<r<\infty$ by a regular HNC
calculation, using a simple iterative procedure. The $g(r)$ for
$r<r_{c}$ is then replaced by the measured $g(r)$, and the process is
repeated until convergence.  For low density, $g(r)$ and $
v(r;\rho_b)$ converge very quickly, but for higher densities, say
$\rho_b/\rho^*>1$, the convergence is slower, and the mixing factor of
the old solution into the new one has to be increased to a value as
large as 99\%.  In fact, because of the finite box-size, the inversion
process is underdetermined, and our ansatz that $ v(r;\rho_b)=0$ for
$r > r_c$ is needed to find a unique solution.  This is not
unreasonable since we don't expect the interactions between the
polymer coils to be significant beyond a distance a few times the
radius of gyration.  However, to make sure that this is actually the
case, we found that relatively large simulation boxes were needed,
with a lattice size of up to $10-15 R_g$.  This is especially
important at high density, where the inverted potential becomes longer
ranged and more sensitive to small changes in the radial distribution
function $g(r)$.  In all our inversions we checked explicitly that $
v(r;\rho_b)$ becomes effectively zero for an $r < r_c$, confirming our
initial ansatz.

\begin{figure}
\begin{center}
\epsfig{figure=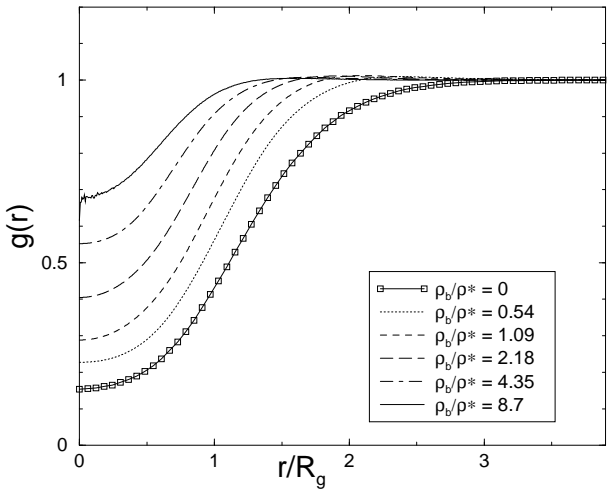,width=8cm}
%\begin{minipage}{8cm}
\caption{\label{fig:gL500} The polymer CM pair distribution function
$g(r)$ calculated for $L$$=$$500$ SAW polymers and used to generate
$  v(r;\rho_b)$.  The x-axis denotes $r/R_g$, where $R_g$ is the
radius of gyration of an isolated SAW polymer.  }
%\end{minipage}
\end{center}
\end{figure}

The resulting radial distribution functions $g(r)$ are shown in
Fig.~\ref{fig:gL500}, and are similar in shape to those of the pure
Gaussian core model\cite{Loui00a}.  As the density goes up the
correlation hole at small $r$ decreases in range and height.  Except
for a small maximum around $r \approx 2 R_g$, the pair correlation
functions do not show oscillations within the statistical noise of
about $0.1\%$, for any density considered here.

\begin{figure}
\begin{center}
\epsfig{figure=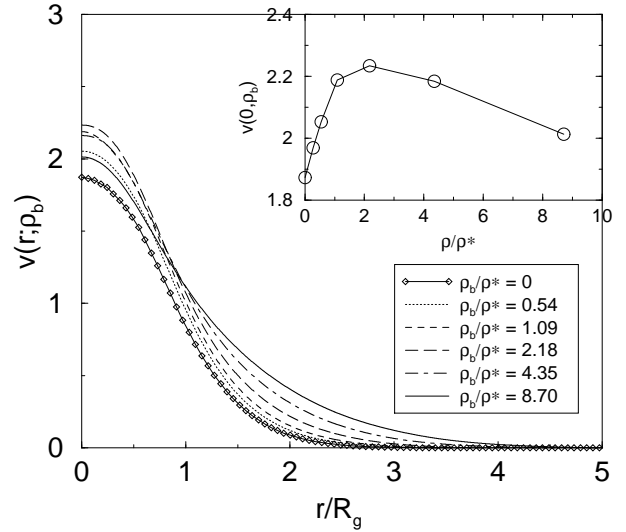,width=8cm}
%\begin{minipage}{8cm}
\caption{\label{fig:veffL500} The effective polymer CM pair potential
$ v(r;\rho_b)$ derived from an HNC inversion of $g(r)$ for different
densities.  The x-axis denotes $r/R_g$, where $R_g$ is the radius of
gyration of an isolated SAW polymer. Inset: The value of the effective
polymer CM pair potential at $r =0$, as a function of density
$\rho_b/\rho^*$. The maximum of the potential initially increases before
decreasing at high concentration.}
%\end{minipage}
\end{center}
\end{figure}

\begin{figure}
\begin{center}
\epsfig{figure=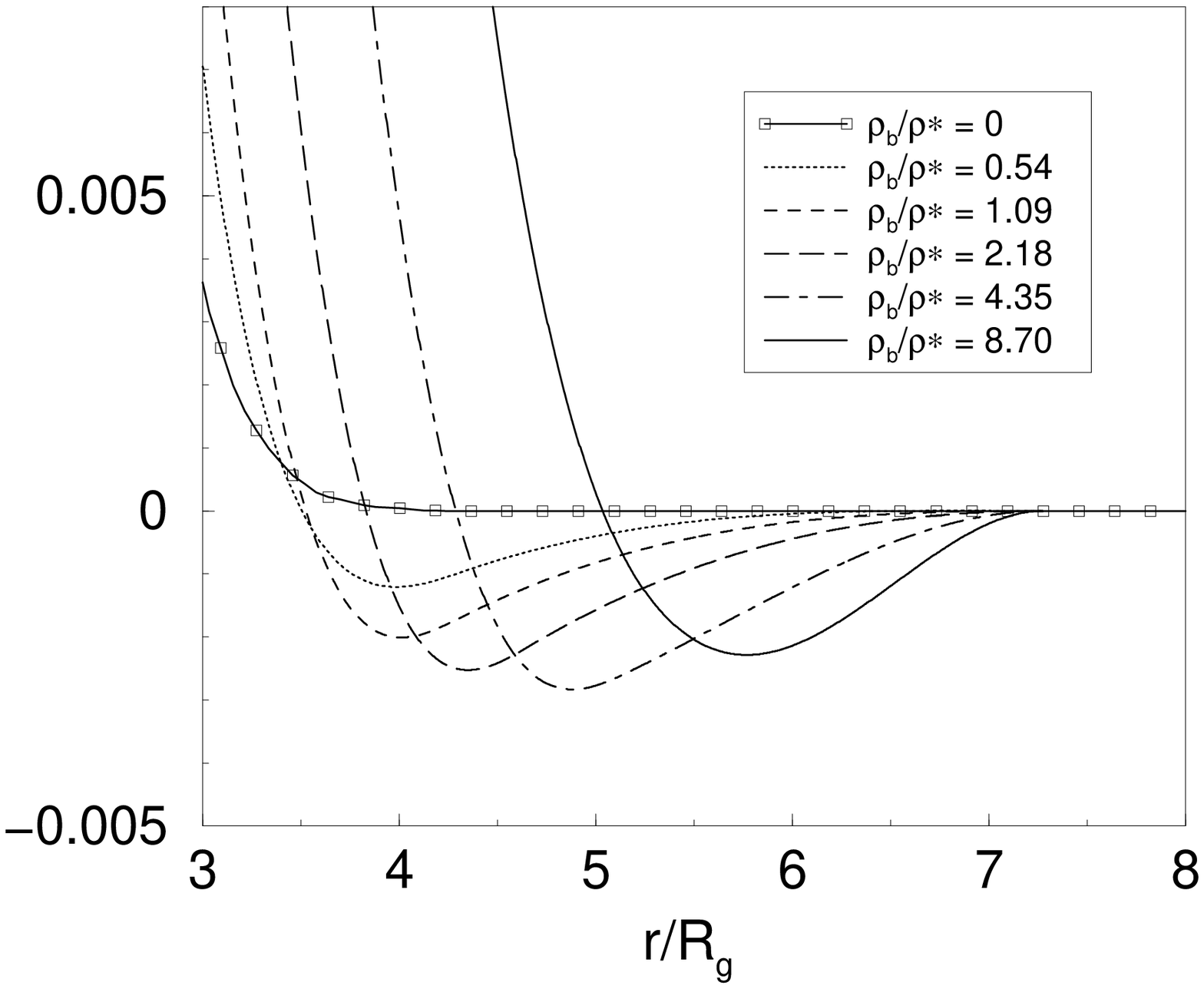,width=8cm}
%\begin{minipage}{8cm}
\caption{\label{fig:veffneg} The negative part of the effective
polymer CM pair potential $  v(r;\rho_b)$ derived from an HNC
inversion of $g(r)$ for different densities.  The x-axis denotes
$r/R_g$, where $R_g$ is the radius of gyration of an isolated SAW
polymer.  }
%\end{minipage}
\end{center}
\end{figure}

The effective polymer-polymer potentials $v(r;\rho_b)$, obtained from
the $g(r)$'s, are shown in Fig.\ref{fig:veffL500}.  Careful inspection
of the figure reveals that the effective pair potential is not very
sensitive to the polymer concentration.  The value at $r=0$ first
increases slightly with $\rho_b$, before decreasing again at the
highest concentrations, as is depicted in the inset of
Fig~\ref{fig:veffL500}, while the range of $ v(r;\rho_b)$ increases
with $\rho_b$.  A more subtle feature, highlighted in
Fig.~\ref{fig:veffneg}, is that the effective potential becomes
slightly negative $({\cal{O}}(10^{-3} k_BT))$ for $r/R_g \gtrsim 3 $
at the higher concentrations.  These effects become apparent only when
large enough box-sizes are used.  Although the negative tails seem
very small, they are nevertheless significant since the thermodynamics
depend on the integral of $r^2 v(r;\rho_b)$.  For example, leaving
them out can easily induce a $5\%$ change in the pressure.  It is,
therefore, paramount to include these effects in (quasi)-analytical
representations of the effective potentials. For that reason, a simple
fit to a Gaussian or a sum of Gaussians is not accurate enough to
reproduce the structure and the thermodynamics of the SAW polymer
systems and consequently, we chose to use an interpolation spline fit
to describe the potentials.  First, the raw effective potential data
were fitted to a Gaussian:
\begin{equation}
v_{est}(r) = a_0 e^{-a_1 r^2}.
\end{equation}
Subsequently, the difference $\Delta v(r;\rho_b) = v(r;\rho_b) - v_{\rm
est}(r)$ was fitted by employing a least squares spline procedure with 8
nodes (the ``dfc'' routine of the slatec library \cite{slatec}).  The
values of the nodes are not known in advance, except for the
boundaries $r=0$ and $r=r_c$. Additional constraints on the spline fit
were: $ v(r_c) =0$, $d v(r=r_c)/dr=0$ and $ v(r=0)/dr=0$.  We
optimized the spline fit by moving the nodes on the x-axis using a
Monte Carlo procedure. The parameters for the fits are available
elsewhere~\cite{web}.

\begin{figure}
\begin{center}
\epsfig{figure=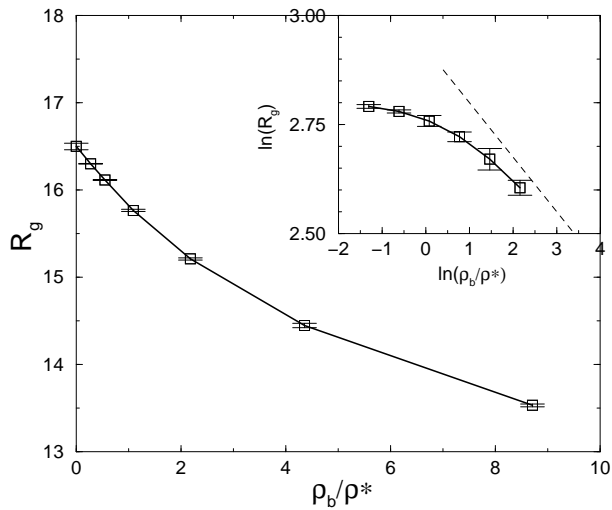,width=8cm}
\caption{\label{fig:r0rho} The effective radius of gyration for
$L=500$ SAW polymers decreases as a function of density $\rho_b/\rho^*$.
{\bf Inset:} At high densities the effective radius of gyration
asymptotically follows the scaling law $R_g \sim \rho^{1/8}$.}
\end{center}
\end{figure}

Note that in Fig.~\ref{fig:veffL500}, the polymer-polymer interaction
$ v(r;\rho_b)$ is plotted v.s.\ $r/R_g$, where $R_g$ is the radius of
gyration of {\em isolated} polymers in the infinitely dilute limit.
In a dense solution, the effective radius of gyration of the polymers
contracts according to the power-law, $R_g \sim
\rho_b^{-1/8}$~\cite{deGe79,Daou75,powerlaw}, as shown in
Fig.~\ref{fig:r0rho}.

\begin{figure}
\begin{center}
\epsfig{figure=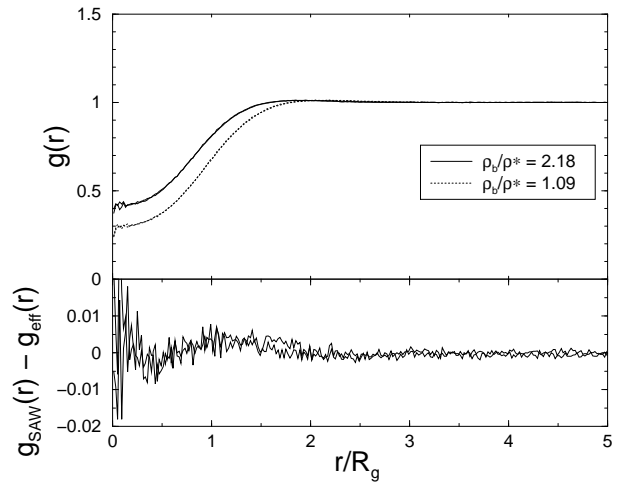,width=8cm}
\caption{\label{fig:checkvpol} The $g(r)$ of a system interacting via
the effective potential $v(r;\rho_b)$ compared with the CM pair distribution
of a SAW simulation for two polymer concentrations. The differences
are shown in the lower panel and are typically less than $\pm 0.01$.  }
\end{center}
\end{figure}

The accuracy of the effective potentials are tested by performing a
 direct Molecular Dynamics (MD) simulation of the ``soft colloids''
 interacting via $v(r;\rho_b)$.  In Fig.~\ref{fig:checkvpol} the pair
 distribution function $g_{\rm MD}(r)$ from MD simulations is compared
 with the original SAW $g(r)$ for two densities in the semi-dilute
 regime. The difference between the two distribution functions shows
 an oscillation at small $r$. Because this occurs in the same way for
 both densities it is possibly introduced by the inversion
 procedure. Even so, the difference between the two distribution
 functions is still typically less then $\pm 0.01$.  We conclude that
 the HNC inversion procedure yields very accurate effective potentials
 for soft particles, capable of describing the structure of the fluid
 with an absolute error of less than $\pm 0.01$.

In a previous paper we have shown that the HNC closure is very
accurate when applied to the Gaussian model\cite{Stil76}, whereby
particles interact via the repulsive potential $v(r)=\epsilon \exp [
-\alpha \left(r/R_g\right)^2]$, and is in fact quasi-exact in the
regime relevant to the effective potentials shown in
Fig.~\ref{fig:veffL500}\cite{Loui00a,Lang00}.  Even the much cruder
RPA closure, $c(r)= - v(r)$, yields semi-quantitatively accurate
results for correlations and thermodynamics in the regime of
interest. Thus polymer solutions in the dilute or semi-dilute regime
fall into the class of mean field fluids according to the nomenclature
introduced in Ref.~\cite{Loui00a}.

The inversion procedure guarantees that the two-body correlations are
accurately reproduced by the effective potential, but this does not
necessarily imply that higher order correlations are also well
represented.  As a first test we performed preliminary simulations of
the three-body bond-order correlation functions $g_3(r,\theta,\varphi)$
for both full SAW walks and our soft particles. The two approaches
lead to identical results within statistical errors, implying that
higher order correlations are much more accurately reproduced than one
might initially expect.  We have also performed some preliminary
calculations of the three-body interaction $v_3({\bf r_1}, {\bf r_2},
{\bf r_3})$. Even at full overlap of the three centers of mass, the
three-body interaction term is only about $10 \%$ of the pairwise
interaction.  This is consistent with the results found for
star-polymers\cite{vonF00}, and was fore-shadowed by the relatively
weak density dependence of the effective pair interaction $v(r;\rho_b)$.

Besides accurately describing the structure, it is also important that
the thermodynamics are captured by the effective potential. In the
next section we therefore focus on the equation of state (e.o.s.)
for polymer solutions.

\subsection{Equation of state}

\label{sec:eos}

\subsubsection{Equation of state from direct SAW simulations}

\begin{figure}
\begin{center}
\epsfig{figure=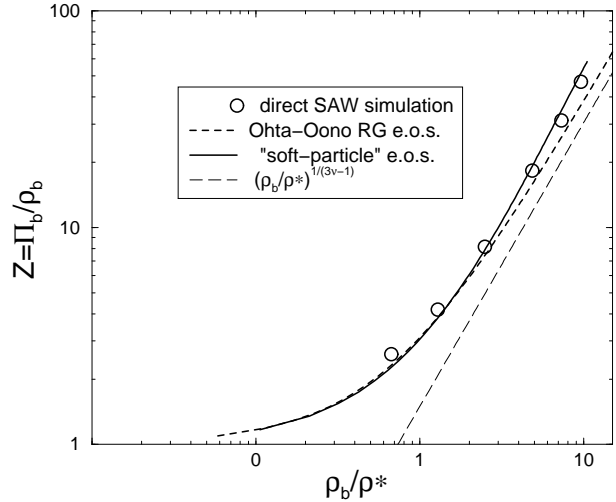,width=8cm}
\caption{\label{fig:Z-log} Log-log plot of the e.o.s.\ $Z= \Pi/\rho_b$
as a function of the density for $L=500$ polymers.  The soft-particle
e.o.s.\ gives a good representation of the full SAW polymer
simulations.  At the highest densities there is a slight deviation
from the expected des Cloizeaux $(\rho_b/\rho^*)^{1/(3\nu-1)}$ scaling
law which we attribute to the effects of a finite monomer
concentration $c$.  Also shown is the RG e.o.s. of Ohta and
Oono\protect\cite{Ohta82,Oono85}.}
\end{center}
\end{figure}

 We measured the e.o.s., $\Pi_b/\rho_b$, directly for a SAW simulation
by using the thermodynamic integration approach of
Dickman~\cite{dickman}. In this method the bulk (osmotic) pressure
$\Pi_b$ is measured by taking the derivative of the free energy $F$
with respect to volume of a system of SAW polymers between two hard
walls. The polymers live on a rectangular cubic lattice of size
$M=H\times D\times D$, which is periodic in the y and z
directions. The two walls are represented by an infinitely repulsive
potential at $x=0$ and at $x=H+1$, so that the polymer segments cannot
penetrate the walls.  The volume of a lattice can only change
discretely, and the free energy derivative changes to a finite
difference
\begin{eqnarray}
  \Pi_b & = & \frac{\partial \ln Z( N,L,D,H)}{\partial M} = D^{-2}
 \frac{\partial \ln Z( N,L,D,H)}{\partial H} \nonumber\\ & \approx &
 D^{-2} \left( \ln Z( N,L,D,H) - \ln Z( N,L,D,H-1) \right).
\end{eqnarray}
The model is modified by associating an additional repulsive potential
$-\ln \lambda$ with each occupied site in the plane $x=H$, where
$0<\lambda <1$.  The partition function then becomes
\begin{equation} 
 Z( N,L,D,H\lambda)= \sum_{\rm polymer \, conf} e^{-  U} \cdot \lambda^{n_H},
\end{equation}
where $n_H = D^2 \rho_H(\lambda)$ is the number of occupied sites in
the $x=H$ plane, and $\rho_H(\lambda)$ is the corresponding number density in this plane. The pressure can now be estimated as
\begin{equation} 
\label{eq:lambdaint}
  \Pi_b = D^{-2} \int_0^1 d\lambda \left( \frac{\partial \ln
Z}{\partial \lambda} \right)= \int_0^1 d\lambda
\frac{\rho_H(\lambda)}{\lambda}
\end{equation}
We performed SAW simulations of polymers with length $L=500$ on a
$M=160\times 100\times 100$ cubic lattice for $N=50, 100, 200, 400,
600$ and $800$.  For each density we determined the value of
$\rho_H(\lambda)$ for 5 different values of $\lambda$, corresponding
to the abscissae of a 5 point Gaussian quadrature which was used to
evaluate the integral in Eq.~(\ref{eq:lambdaint}) The resulting e.o.s.\
is plotted in Fig.~\ref{fig:Z-log}.

\subsubsection{Equation of state from the  soft-particle picture}

To calculate the e.o.s.\ within the soft-particle picture, we use the
compressibility relation (\ref{eq4.2}), which must now be integrated
w.r.t.\ the density:
\begin{equation}\label{eq5.1}
  \Pi_b(\rho_b) = \int_0^{\rho_b} (1- \rho^\prime \hat{c}(0,\rho^\prime))
d\rho^\prime.
\end{equation}
We used the quasi-exact HNC approximation to calculate $c(r)$ from the
inverted effective potential $ v(r;\rho_b)$ for several state-points,
fitted the values of $\hat{c}(0;\rho_b)$, and integrated w.r.t.\
density.  As demonstrated in Fig.~\ref{fig:Z-log}, the e.o.s.\ is very
close to the one obtained by direct SAW simulations, immediately
suggesting that our inversion procedure indeed reproduces the true
thermodynamics of the full many-body system.  This success also
implies that the volume terms are small, possibly smaller than the
statistical error in our present simulations and inverted
potentials. In fact, using a simple scaling theory, Likos has argued
that the contribution of volume terms to the e.o.s.\ scales as
$(\rho_b/\rho^*)^{3/8}$ in the semi-dilute regime, and so contributes
little to the full e.o.s.\cite{Liko00}.

Also shown in Fig.~\ref{fig:Z-log} is the RG result by Ohta and
Oono~\cite{Ohta82}; we use a slight improvement with correct
exponents\cite{Oono85}. The one remaining fit parameter is determined
by the second osmotic virial coefficient $B_2$ for $L=500$ SAW
polymers, a procedure similar to that used when comparing to
experiment\cite{Wilt83}.  The agreement is seen to be fairly good,
although the SAW e.o.s.\ is somewhat higher than the RG results. This
is most likely due to the fact that the monomer density $c$ is not
zero, which induces small corrections to the full scaling limit (see
the discussion in Section \ref{sec:polpolinv}).

\begin{figure}
\begin{center}
\epsfig{figure=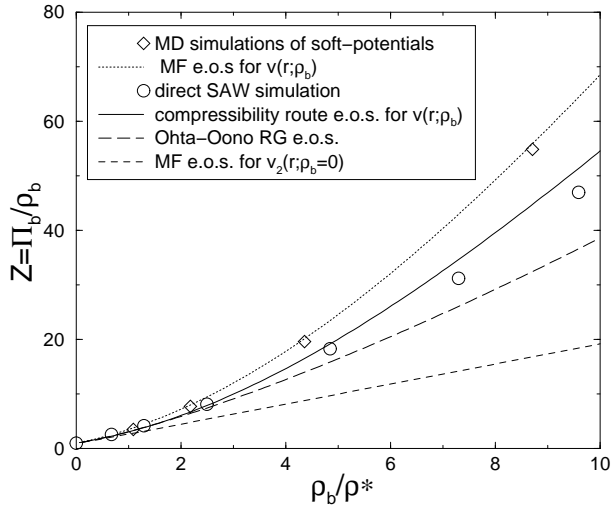,width=8cm}
\caption{\label{fig:Z-linear} Linear plot of the the e.o.s.\
 $Z= \Pi_b/\rho_b$ as a function of the density for $L=500$
polymers.  Several approximations to the e.o.s., discussed in the text, are
compared.}
\end{center}
\end{figure}

Instead of the compressibility route, one could also use the virial
route to the e.o.s.\cite{Hans86}:
\begin{eqnarray}
\label{eq:densdepvirial}
\frac{\Pi_b}{\rho_b} & = & 1 + \frac{d V_0(\rho_b)}{d \rho_b} \nonumber \\
& - & \frac{1}{3} \sum_{i < j}^N \left< r_{ij} \frac{\partial
v(r_{ij};\rho_b)}{\partial r_{ij}} - 3 \rho_b \frac{\partial
v(r_{ij};\rho_b)}{\partial \rho_b} \right>,
\end{eqnarray}
which includes not only the density dependence of the effective pair
potential, but also the density dependence of the volume term.  The
full density dependence of the potentials is at present hard to
calculate, so instead we initially ignore the density derivative and
the volume terms. First, we directly measured the e.o.s.\ of the
soft-particle fluid by a MD simulation with the spline-fit potentials.
The pressure follows from the usual virial theorem, when the density
derivatives in Eq.~(\ref{eq:densdepvirial}) are neglected.  The e.o.s.\
from this approach is depicted in Fig.~\ref{fig:Z-linear}, and
compared to the simple mean-field (MF) form:
\begin{equation}
Z_{MF} = 1 + \frac{1}{2}   \hat{v}(0;\rho_b)\rho,
\end{equation}
which gives a good fit to the simulations, as expected for soft-core
fluids in the MFF regime\cite{Loui00a}. Here $\hat{v}(0;\rho_b)$ is the
$k=0$ component of the FT of the pair interaction.  However, by
including only the explicit density dependence of the effective pair
potentials while ignoring the density derivative terms in the virial
equation, we {\em overestimate} the e.o.s.\ compared to the full SAW
simulation.

The density dependence can be neglected even further by simply taking
the $\rho_b \rightarrow 0$ form of the pair potential, $ v_2(r)$, and
applying it at all densities.  The resulting e.o.s.\ now {\em
underestimates} the e.o.s.\ when compared to the full SAW simulation,
as demonstrated in Fig.~\ref{fig:Z-linear}.  We note that a similar
approach was employed in recent work on the phase-behavior of
star-polymers, where the $\rho_b\rightarrow 0$ limit of the pair
potential was used to calculate the structure and phase-behavior at
finite concentration\cite{Watz99}.

Finally, we comment on the common practice of extracting osmotic
virial coefficients from the measured experimental e.o.s.  Firstly,
the virial equation has a very small radius of convergence for
soft-core fluids\cite{Loui00a}. Secondly, the range of the effective
potential $v(r;\rho_b)$ increases with density.  These two effects imply
that a naive linear fit to all but the very lowest polymer
densities will lead to an overestimate of the true osmotic second
virial coefficient $B_2$.

\section{Effective Wall-Polymer potentials}

\subsection{Polymer coils near a wall}

Polymer coils near a non-adsorbing hard wall exhibit a depletion layer
due to entropic effects.  This is true even for ideal Gaussian
polymers, and if one were to model these by effective CM potentials,
the polymer-polymer potential would be zero, but there would still be
a polymer-wall potential of the form $ \phi(z) = \ln(\rho(z)/\rho_b)$,
where $\rho(z)$ is the CM density profile near the wall and $\rho_b$
is the uniform density far from the wall (see the Appendix for more
details).  Thus a complete description of polymer coils in confined
geometries requires not only the polymer-polymer interactions derived
in the previous section, but also effective polymer-wall potentials $
\phi(z;\rho_b)$.

\begin{figure}
\begin{center}
\epsfig{figure=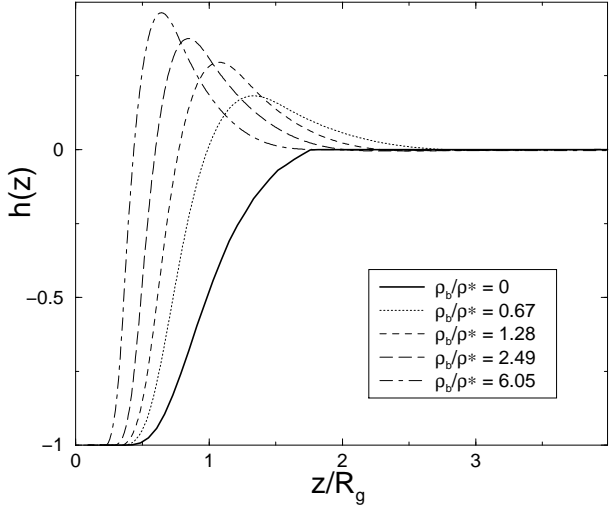,width=8cm}
%\begin{minipage}{8cm}
\caption{\label{fig:hz} The wall-polymer CM density profile $h(z) =
\rho(z)/\rho_b-1$ for SAW polymers at different bulk concentrations.
From $h(z)$ we can calculate the corresponding polymer adsorptions
$\Gamma$ and find $-\Gamma = 0, 0.094, 0.13, 0.16$, and $0.20$ in
units of $R_g^{-2}$ respectively. The relative adsorptions are
$-\Gamma/\rho_b = 0.84, 0.59, 0.41, 0.27$, and , $0.14$
respectively, and decrease with increasing density as expected.  }
%\end{minipage}
\end{center}
\end{figure}

 We follow a strategy similar to that used in the
homogeneous case, and first calculate the wall-polymer density profile
$\rho(z)$, from which we then extract an effective potential $
\phi(z;\rho_b)$.  Using the same explicit SAW polymer model as in
Section~\ref{sec:model}, we performed MC simulations of polymers of length
$L=500$ on a lattice of size $M=160\times 100\times 100$ with hard
walls at $x=0$ and $x=160$. The polymer segments were not allowed to
penetrate the walls. The simulations were done for $N=50, 100, 200$
and $500$. During each simulation, we computed the density profiles
$\rho(z)$, where $z$ denotes the distance of the polymer CM from the
wall and $\rho_b$ is the bulk density far from the wall.  The normalized
profiles $h(z)=\rho(z)/\rho_b -1$, for different bulk concentrations
$\rho_b/\rho*$ are shown in Fig.~\ref{fig:hz}.

 The polymer coil adsorption $\Gamma$ is defined by:
\begin{equation}\label{eq:adsorption}
\Gamma = -\frac{\partial( \Omega^{ex}/A)}{\partial \mu} = \rho_b
\int_0^{\infty} h(z) dz,
\end{equation}
where $\Omega^{ex}/A$ is the excess grand potential per unit area and
 $\mu$ the chemical potential of the polymers.  As the density
 increases, more polymer is adsorbed at the wall as expected, but the
 relative adsorption, $\Gamma/\rho_b$, decreases.

\begin{figure}
\begin{center}
\epsfig{figure=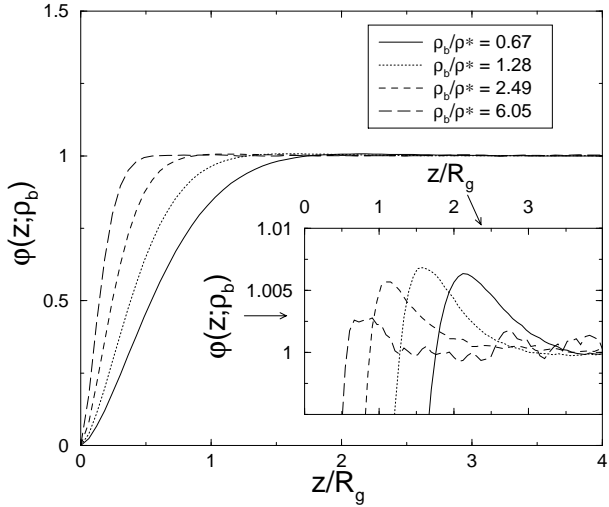,width=8cm}
%\begin{minipage}{8cm}
\caption{\label{fig:monomer} The wall-monomer density profile
$\varphi(z;\rho_b) = \rho(z)/\rho_b-1$ for the same set of densities 
as in Fig.\protect\ref{fig:hz}. {\bf Inset:} A magnification of 
 the
region where there is a small correlational bump  in the density
profiles.  The height is less than $1\%$ of the total density while the
range is about  $ R_g$, implying that the bump  arises from
polymer-polymer correlations.}
%\end{minipage}
\end{center}
\end{figure}

The normalized monomer density profiles for SAW's are shown in
Fig.~\ref{fig:monomer} for the same polymer densities as the CM
profiles shown in Fig.~\ref{fig:hz}.  As expected, the profile moves
closer to the wall for higher density; the width of the monomer
depletion layer shifts from around $R_g$ at the lowest densities, down
to values dictated by the segment correlation length~\cite{deGe79} in
the semi-dilute regime. Although the profiles do not show such a clear
correlation-induced oscillation as the CM profiles, there is
nevertheless still a small maximum in the depletion layer as
illustrated in the inset of Fig~\ref{fig:monomer}.  The peak in the
monomer profile is less than $1\%$ of the bulk density, and seems to
decrease for higher overall polymer concentration. The range is about
$R_g$, implying that it arises from correlations between polymer
coils.  We observe only one peak, although due to statistical noise,
we cannot rule out the possibility of more oscillations in the density
profiles.  Recently self-consistent field calculations, valid for
polymers in a theta solvent, found a similar small oscillation in the
monomer profiles\cite{Guch00}.

\subsection{Deriving $  \phi(z;\rho_b)$ from $\rho(z)$}
\label{wall-polymer}

  From a knowledge of the concentration profile $\rho(z)$, and the
bulk direct correlation function between polymer CM's, $c_b(r)$, one
may extract an effective wall-polymer potential $\phi(z;\rho_b)$ by combining
the wall-polymer OZ relations\cite{Hans86} with the HNC closure.
For a binary mixture of two components labeled 0 and 1, in which
component 0 is infinitely dilute $(x_0 \rightarrow 0)$, the
Ornstein-Zernike equations become~\cite{Hans86}
\begin{mathletters}
\label{eq6.1}
\begin{eqnarray}
h_{11}(1,2) & = & c_{11}(1,2) + \rho_b \int h_{11}(1,3) c_{11}(2,3) d3
\label{eq6.1a}\\
h_{10}(1,2) & = & c_{10}(1,2) + \rho_b \int h_{11}(1,3) c_{10}(2,3) d3
\label{eq6.1b}\\
h_{01}(1,2) & = & c_{01}(1,2) + \rho_b \int h_{01}(1,3) c_{11}(2,3) d3
\label{eq6.1c}\\
h_{00}(1,2) & = & c_{00}(1,2) + \rho_b \int h_{01}(1,3) c_{10}(2,3) d3
\label{eq6.1d}
\end{eqnarray}
\end{mathletters}
In the limit $R_0 \rightarrow
\infty$, Eq.~(\ref{eq6.1c}) becomes an equation for the wall-density
profile, sometimes called the wall-OZ relation:
\begin{equation}\label{eq:walloz}
h(z) = c_{01}(z) + \rho_b \int d{\bf r'} h_{01}(z') c_b(|{\bf r
- r'}|),
\end{equation}
where $h(z) = \rho(z)/\rho_b -1$. The wall-OZ equation can be solved, given the bulk correlation function $c_b(r)$, and
a closure relation.  In Section.~\ref{sec:polpolinv} we showed that the HNC
closure gives excellent results for effective polymer-polymer
interactions, and it is therefore natural to apply the same
approximation here. Combining Eq.~(\ref{eq:hnc}) with
Eq.~(\ref{eq:walloz}) we obtain
\begin{equation}\label{eq:stells}
  \phi(z;\rho_b) =   \phi^{MF}(z;\rho_b) + \rho_b \int d{\bf r'} h(z')
c_b(|{\bf r} - {\bf r'}|).
\end{equation}
\begin{figure}
\begin{center}
\epsfig{figure=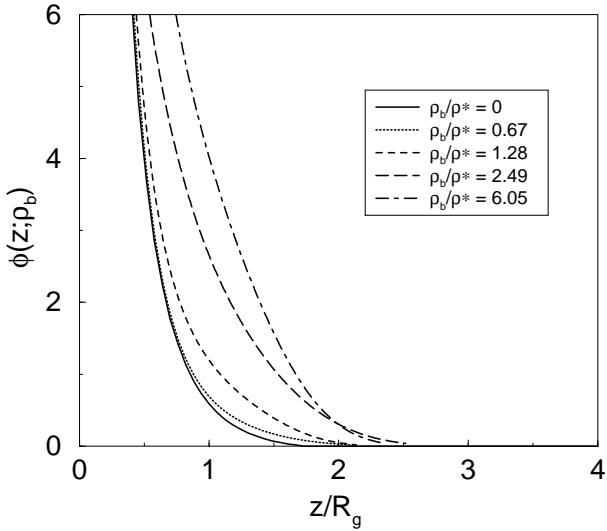,width=8cm}
%\begin{minipage}{8cm}
\caption{\label{fig:vz} The wall-polymer potential $\phi(z;\rho_b)$ as
obtained from the inversion of $h(z)$ via the HNC expression, Eq.~
(\protect\ref{eq:stells}).  }
%\end{minipage}
\end{center}
\end{figure}
\noindent The first term is the usual potential of mean force $
\phi^{MF}(z;\rho_b) = - \ln \left[\rho(z)/\rho_b\right]$, to which $
\phi(z;\rho_b)$ would reduce in the $\rho_b \rightarrow 0$ limit,
while the second term arises from correlations between the polymer
coils next to the wall.  An identical equation results from the HNC
density functional theory (DFT) approach\cite{Evan92}, and a similar
one, with $c_b(r)$ replaced by $ v(r;\rho_b)$ obtains if a mean field
DFT is used.  In contrast to simple fluids, where
Eq.~(\ref{eq:stells}) is not very reliable, the wall-HNC closure works
remarkably well for the Gaussian core-fluid in the regime relevant to
polymer solutions\cite{Loui00a}.  Using the $c_b(r)$ extracted from
the earlier bulk simulations of $g(r)$ (see Section~\ref{sec:polpolinv}),
together with Eq.~(\ref{eq:stells}), we are able to extract $
\phi(z;\rho_b)$ from the density profiles.  In order to calculate the
integral in Eq.~(\ref{eq:stells}), we use the procedure outlined by
Sullivan and Stell~\cite{stell}. In contrast to the inversion of the
bulk $g(r)$, where we had to iterate until convergence, the
wall-polymer inversion requires only one step since $c_b(r)$ is given
once and for all.  Results for various bulk concentrations are plotted
in Fig.~\ref{fig:vz}.  The range of the effective wall-polymer
repulsion increases with increasing concentration, while the density
profiles actually move in closer to the wall.  The compression and
enhanced correlation in the density profiles with increasing density
resembles that of the pure Gaussian core fluid in a fixed external
potential\cite{Loui00a}, but the effect is less pronounced in the
former case since for polymer solutions the wall-polymer potential
becomes more repulsive with density. This is due mainly to the
correlation term, which is nearly linear in $\rho_b$, and so becomes
relatively more important as the density increases.  Nevertheless at
shorter distances the $\phi^{MF}(z;\rho_b)$ term still dominates.  The
importance of including both the potential of mean force, and the
correlation-induced component of the effective potential is
demonstrated in Fig,~\ref{fig:ib}.  At very low densities the
potential of mean force is adequate, but at higher densities the
correlation term becomes increasingly important.

\begin{figure}
\begin{center}
\epsfig{figure=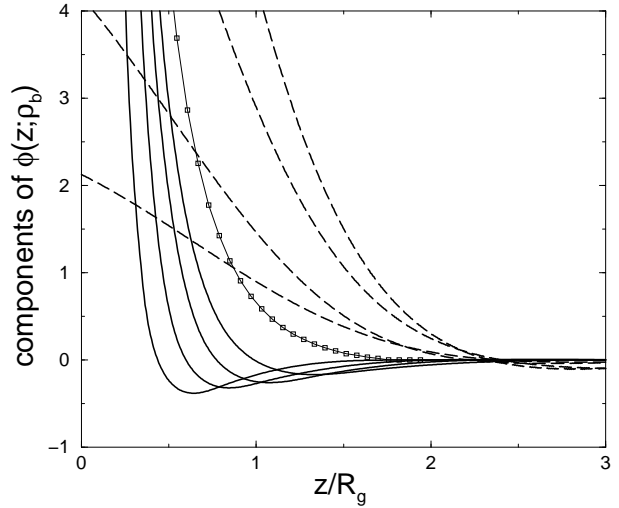,width=8cm}
\caption{\label{fig:ib} Comparison between the contributions to the
effective wall-polymer potentials from the potential of mean force
(solid lines) and from the correlation part (dashed lines) (cf.
Eq.~(\protect\ref{eq:stells})) for polymer concentrations $\rho_b/\rho^*=
0.67, 1.28, 2.49$ and $6.05$. From top to bottom the solid lines
correspond to increasing density and the dashed lines correspond to
decreasing density. The solid line with the small squares denotes the
potential of mean force for infinitely diluted systems. }
\end{center}
\end{figure}

The effective potentials decay exponentially, and to obtain a useful
analytic form for the effective potential, the logarithm of $
\phi(z;\rho_b)$ can be fitted to a cubic polynomial, which describes the
potential very well. However, as in the bulk case, the wall-polymer
potential $\phi(z;\rho_b)$ has a small negative component that cannot
be described by an exponential function. Although in this case the
tail is probably not very important, in order to be consistent, we fit
$ \phi(z;\rho_b)$ by a least squares spline fit similar to the one
described Section.~\ref{sec:polpolinv}.  The parameters for this fit are
available elsewhere~\cite{web}.

\subsection{Consistency of the wall-polymer inversion}

\begin{figure}
\begin{center}
\epsfig{figure=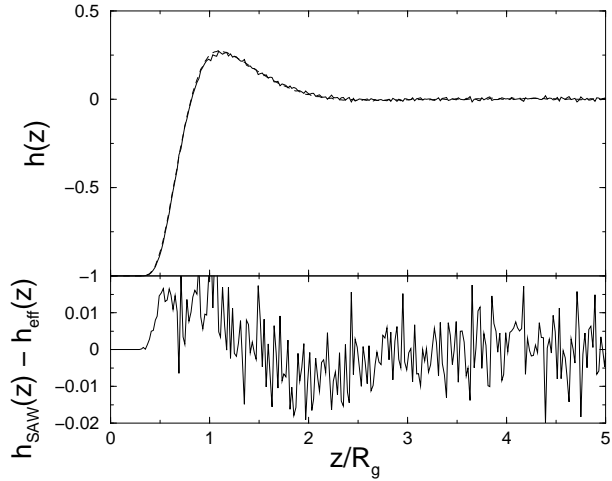,width=8cm}
\caption{\label{fig:checkvcolpol} The profile $h(z)$ of a system of soft
colloids near a wall (dashed line). The particles interact with each
other via $v(r)$, and with the wall via $\phi(z;\rho_b)$. This is
compared with the wall-CM distribution of an explicit SAW simulation
(solid line). The difference is shown in the lower panel and is less
than $\pm 0.02$.  }
\end{center}
\end{figure}

To test the validity of the inversion procedure for the wall-polymer
 $\rho(z)$, we performed Molecular Dynamics simulations of a system of
 ``soft colloidal'' particles interacting with each other via the
 effective pair potential $v(r,\rho_b)$ and with a wall via the
 inverted potential $ \phi(z;\rho_b)$ for the appropriate bulk
 concentration $\rho_b$.  Such effective potential simulations are at
 least an order of magnitude faster than simulations of the original
 SAW model.  The resulting concentration profile of the effective
 particles  is shown in Fig.~\ref{fig:checkvcolpol} for one
 density; it agrees to within an absolute error of roughly $\pm 0.02$ with
 the $\rho(z)/\rho_b$ obtained from the detailed SAW
 simulations.  The corresponding adsorption $\Gamma$ also differs by
 less than $1 \%$ from the value obtained by the SAW simulation, thus
 demonstrating the adequacy of the soft colloid representation of
 the interacting polymer coils, and the accuracy of the HNC inversion
for polymer coils near a hard wall.

\section{Depletion potential between two walls}

\subsection{Full SAW simulations}

One of the aims of this work is to show that the soft particle
description of the polymers provides a useful route to the
colloid-colloid depletion potential in mixtures of colloidal particles
and non-adsorbing polymers. Calculating these depletion interactions
poses a severe test of the soft colloid representation.

 As a first step we calculate the depletion potential between two
planar walls, which can ultimately be applied to spherical colloids through
the Derjaguin approximation.  We confined the polymers within a slit
of width $d$, and, using direct grand-canonical simulations of
the full SAW polymer model, we computed the osmotic pressure exerted
by the polymer coils on the walls.  The insertion of polymers was
achieved by the configurational bias Monte Carlo
technique~\cite{frenkelbook}. The (osmotic) pressure $\Pi(d)$ was
calculated for different values of the spacing $d$ between the walls
by a thermodynamic integration technique similar to the one explained
in Section~\ref{sec:eos}.  Details of these simulations can be found in
Ref.~\cite{Meij00}.  The interaction free energy per unit area A,
$  \Delta F/A$, is then obtained by integrating the osmotic
pressure as a function of $d$:
\begin{equation}
\label{eq:intP}
\Delta F(d)/A = \int_d^\infty dz (\Pi(z) - \Pi(\infty) ),
\end{equation}
where $\Pi(\infty)$ denotes the bulk osmotic pressure $\Pi_b$.  These
 explicit SAW simulations are rather computer intensive, and were
 only carried out for $L=100$\cite{Meij00}.

\subsection{Effective potential simulations}

 In the soft colloid picture, the interactions of the polymer CM's
with each other, $v(r;\rho_b)$, and with a wall, $\phi(z;\rho_b)$, are
calculated once with the HNC inversion procedures from the $g(r)$ and
$\rho(z)$ of a full SAW polymer simulation at the bulk density
$\rho_b$.  These potentials are then used in grand-canonical MC
simulations of soft particles between two walls. The imposed chemical
potential is chosen such that for infinitely separated walls the bulk
density is recovered. The (osmotic) virial pressure is measured as a
function of wall separation $d$, and the interaction free energy per
unit area $  \Delta F/A$, is again obtained by integration of the
pressure via Eq.~(\ref{eq:intP}).

\begin{figure}
\begin{center}
\epsfig{figure=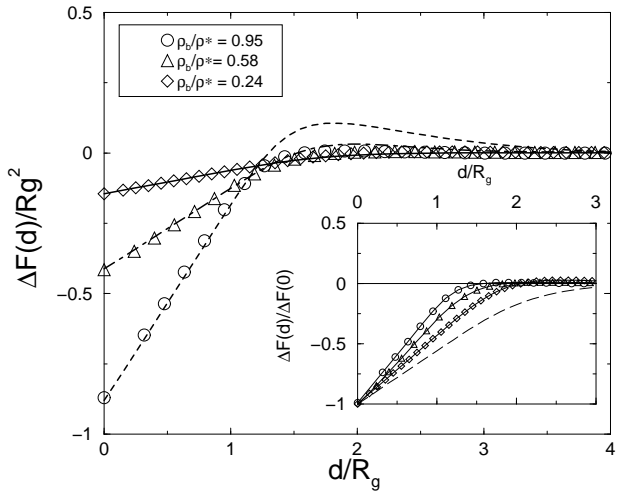,width=8cm}
\caption{\label{fig:vcolcol_rho} Depletion free-energy $ \Delta
F(d)/R_g^2$ between two plates separated by $d$, for three densities,
$\rho_b/\rho^* = 0.95$, $\rho_b/\rho^* = 0.58$, $\rho_b/\rho^* =
0.24$.  The symbols denote the ``exact'' MC simulations, while the
dashed, dash-dotted and solid lines are the soft-colloid simulations
for the same densities.  Inset: $\Delta F(d)/\Delta F(0)$ for the SAW
simulations, the solid lines are to guide the eye.  The long-dashed
line is the ideal Gaussian polymer result calculated in the Appendix.
Note that the range decreases with density, and that, even for the
lowest density, the AO ideal polymer approximation overestimates the
interaction range.}
\end{center}
\end{figure}

In Fig.~\ref{fig:vcolcol_rho} the soft colloid depletion
interaction is compared to that of the ``exact'' grand-canonical MC
simulations of $L=100$ SAW polymers, for three different densities,
$\rho_b/\rho^* = 0.28, 0.58$ and $0.95$.  The two approaches are in
good agreement, but the soft colloid calculations are at least two
orders of magnitude faster than the SAW simulations. As expected, the
depth of the potential increases, whereas the range of
the interaction decreases as the density increases\cite{Joan79}.  At the two
lowest densities, the two approaches agree very well, but for
$\rho_b/\rho^* = 0.95$ they differ slightly around $z=2 R_g$ where the
soft particle picture shows a larger repulsive barrier. The barrier
height is, however, small compared to the attractive minimum at
contact, which agrees well with the ``exact'' data, as does the slope
of the attraction.

\begin{figure}
\begin{center}
\epsfig{figure=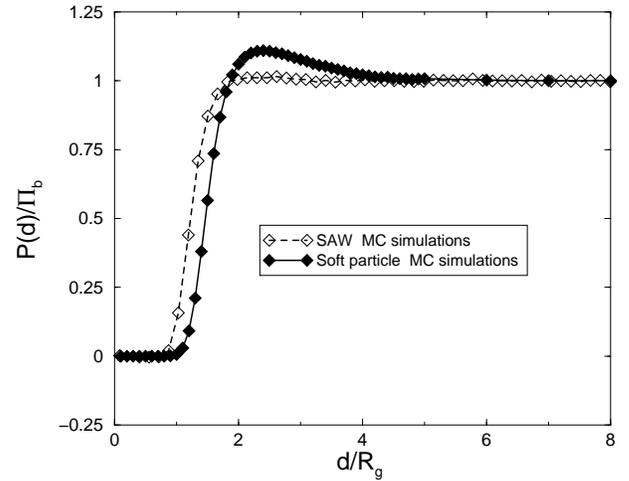,width=8cm}
%\begin{minipage}{8cm}
\caption{\label{fig:Pprofile} Comparison of normalized the depletion
force per unit area (pressure) $P(d)/\Pi_b$ for SAW polymers and soft
particles between two plates separated by $d$ for a density 
$\rho_b/\rho^* = 0.95$.  }
%\end{minipage}
\end{center}
\end{figure}

Liquid state theories for fluids with repulsive particle-particle
interactions predict a repulsive barrier\cite{Gotz99}, so it is not
surprising that the soft particle picture shows a small repulsive
barrier as well. Instead, it is the lack of a significant barrier for
the pure SAW polymer simulations which requires explanation. We trace
this effect to the breakdown of the ``potential overlap
approximation'' (POA) described in the Appendix. Under close
confinement, the interaction of the soft particles with two parallel
walls a distance $d$ apart can no longer be written as the sum of the
two individual wall-particle interactions as would be the case for
simple liquids.  This is caused mainly by the deformation of the
polymers due to the two walls, and also holds for ideal polymers.  The
failure of the POA can be clearly seen in Fig.~\ref{fig:Pprofile},
where we compare the pressure (or force) profiles for the SAW
calculations and the effective potentials.  In the soft particle
picture the pressure starts to rise at a larger inter-wall distance
than the pressure for the SAW polymers, an effect also seen when
non-interaction polymers are represented by an effective particle
representation based on the CM (see the Appendix).  Note that the over
and under-estimates of the pressure cancel each other, so that the
free energy at contact, $\Delta F(0)$, for the effective potentials is
in good agreement with the SAW calculations.

\begin{figure}
\begin{center}
\epsfig{figure=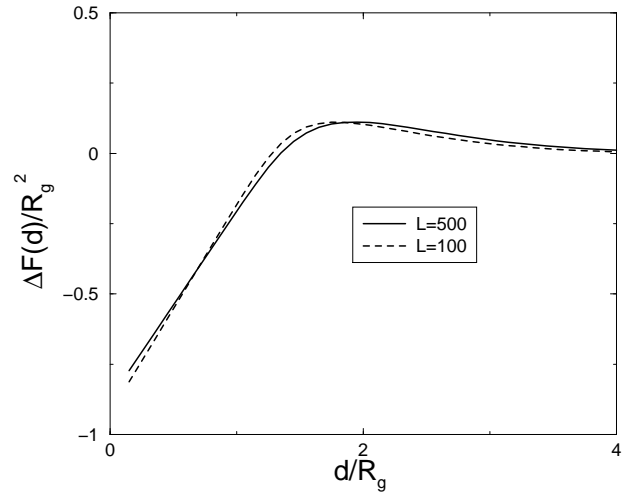,width=8cm}
\caption{\label{fig:comp100/500} Depletion free-energy $ \Delta
F(d)/R_g^2$ between two plates separated by $d$ based on the soft
particle representation for polymers of length $L=100$ and $L=500$.
Here $\rho_b/\rho^* = 0.95$.  }
%\end{minipage}
\end{center}
\end{figure}

The MC simulations for the soft colloid model were carried out
with effective wall-polymer and polymer-polymer potentials appropriate
for $L=100$, since longer polymers are not easily handled in the full
SAW model.  However, we checked that the data obtained with effective
interactions appropriate for longer polymers ($L=500$), are very close
to the $L=100$ results, as is shown in
Fig.~\ref{fig:comp100/500}. Therefore, we are confident that we are
close enough to the scaling regime for the properties under
consideration.

\subsection{Comparison with the Asakura-Oosawa Approach}

The first (and still most popular) approach to the depletion
interaction in colloid-polymer mixtures was pioneered by Asakura and
Oosawa in 1954\cite{Asak54}, when they approximated the polymers as
ideal (Gaussian), and calculated the induced attraction between two
walls.  We shall refer to this neglect of polymer-polymer repulsion as
the {\em AO approximation}, in contrast to the {\em AO model}, where a
further step is taken and the polymers are approximated as
inter-penetrating spheres of radius $R_g$\cite{Asak58}. 

\begin{figure}
\begin{center}
\epsfig{figure=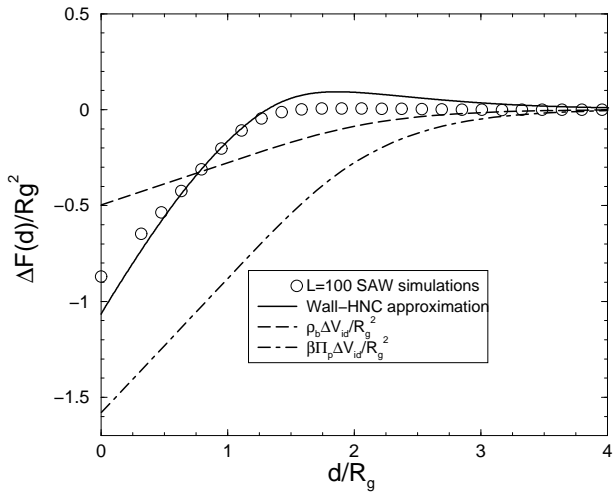,width=8cm,angle=0}
\caption{\label{Fig3} Depletion free-energy $ \Delta F(d)/R_g^2$
between two plates separated by $d$ for $\rho_b/\rho^* = 0.95$ .
Circles are the ``exact'' MC simulations of SAW polymers. The
long-dashed and dash-dotted lines denote the two AO approximations
mentioned in the text.  The short-dashed
line denotes the more accurate wall-HNC approximation of
Eq.~(\protect\ref{eq4}), which is in fact very close to the the
simulations in the soft particle picture shown in
Fig.~\protect\ref{fig:vcolcol_rho}.  }
\end{center}
\end{figure}

The exact depletion potential induced by ideal polymers between two
 plates of area $A$ a distance $d$ apart is given by:
\begin{equation}\label{AO1}
\Delta F(d)/A = \rho_b \Delta V_{id}(d),
\end{equation}
where $\Delta V_{id}(d)$ is the gain in volume accessible to an 
ideal  Gaussian
polymer of size $R_g$, due to overlap of the exclusion volumes close
to the plates.  This can be exactly calculated as shown in the
Appendix. To treat interacting polymers, a widely used
phenomenological improvement (see for example ref.~\cite{Ilet95})
replaces the ideal polymer density by the bulk osmotic pressure
$\Pi_b$ of the interacting polymers in the left over free-volume:
\begin{equation}\label{AO2}
\Delta F(d)/A = \Pi_b \Delta V_{id}(d).
\end{equation}
In Fig.~\ref{Fig3} we plot these two versions of the AO approximation
for the largest density considered above, $\rho_b/\rho^* =0.95$, and
compare them to the effective potential and ``exact'' SAW simulation
results.  The two approaches result in rather poor representations of
both the depth and the range of the true potential, even though we are
technically not yet into the semi-dilute regime where one might expect
them to break down (see also the inset of Fig.~\ref{fig:vcolcol_rho}).
For the lower densities the AO approximation works somewhat better, as
expected.

\subsection{HNC Wall-Wall Approximation}

Following arguments similar to those used to derive the wall-polymer
HNC equations of section \ref{wall-polymer}, one can also derive an
HNC type equation for the depletion interaction free energy per unit
area between two walls separated by a distance $d$\cite{Atta91}:
\begin{eqnarray}\label{eq4}
\frac{  \Delta F(d)}{A} & = & - \rho_b \int_{-\infty}^\infty h(s)
h(d -s) ds \nonumber \\ & +& \rho_b \int_{-\infty}^\infty h(d -s) \left
[   \phi(s;\rho_b) -   \phi^{MF}(s;\rho_b) \right] ds.
\end{eqnarray}
Here $h(s)= \rho(s)/\rho_b -1$ is the {\em single wall} density
profile, $\phi(s;\rho_b)$ is the corresponding effective wall-polymer
potential, and $ \phi^{MF}(s;\rho_b)$ is the corresponding potential of mean
force.  The first term on the r.h.s.\ of Eq.~(\ref{eq4}) represents
the density overlap approximation discussed in the Appendix, and is
the only contribution in the case of ideal Gaussian polymer coils.
The second term arises from correlations between the polymer coils,
and dominates the first term for larger densities.  Note that only
information from one single wall enters into this HNC wall-wall
approach.  We use the effective wall-polymer potential $\phi(z;\rho_b)$ and
the related density profile $h(z)$ from the soft-colloid picture
together with Eq.~(\ref{eq4}) to derive the HNC wall-wall depletion free
energy. As shown in Fig~\ref{Fig3}, this compares well with the MC
simulations of the soft particles except at short distances where a
small deviation develops that can be traced to the fact that only
information from a single wall is used.  A more promising approach,
without this shortcoming, would be to directly use the MF or HNC DFT
approaches applied in ref.~\cite{Loui00a} to Gaussian-core
potentials.

\section{Discussion and Conclusion}
The coarse-grained representation of polymer coils as soft
colloids, put forward in this paper, has proved very reliable. The
effective polymer-polymer and wall-polymer interactions obtained by a
systematic inversion procedure based on fluid integral equations,
yield pair distribution functions and concentration profiles which
agree closely with the results from simulation of the full SAW segment
model, while allowing a massive reduction in computer time compared to
the lattice simulations. Much of the success of the present
coarse-graining procedure lies in our finding that the optimum
effective pair potential between the CM's of neighboring coils does
not depend strongly on polymer concentration, and is reasonably close
to its infinite dilution limit. The effective polymer-polymer and
wall-polymer interaction lead to a rather accurate description of the
depletion interaction between two hard walls, despite the implicit
potential superposition assumption and the fact that the
coarse-graining procedure in its present form does not allow for the
deformation of the polymer coils, away from the spherical shape, in the
vicinity of an impenetrable surface. Such shape fluctuations are
allowed in the alternative procedure by Murat and Kremer~\cite{Mura98}
but the remarkable agreement between the original full SAW model and
the coarse-grained model illustrated in Fig.~\ref{fig:vcolcol_rho}
seems to indicate that shape deformation of confined polymers may not
be a crucial factor to reproduce concentration profiles.

The present inversion procedure yields concentration dependent
effective pair potentials, but does not provide direct access to the
internal free energy of polymer coils~\cite{Mura98}, which plays a
role rather similar to that of the ``self energy'' or volume term of
electric double-layers in charge-stabilized colloidal dispersions or
solutions of star polymers~\cite{Graf98}. This
concentration-dependent term contributes to the osmotic equation of
state, but the good agreement between full SAW model simulations and
the results based on the effective pair interaction without the
volume term would indicate that the concentration dependence
of the latter is weak.

Finally, it must be stressed that the present inversion procedure is
by no means restricted to the simple SAW model of non-intersecting
polymers.  We are in fact planning to extend the coarse-graining
procedure to the case where the segment-segment coupling has an
attractive component to describe the situation of polymer coils in
poor solvent. The case of semi-dilute solutions of polymers of
different lengths will also be considered within the same theoretical
framework with the objective of studying possible demixing, as
suggested by our recent investigation of binary Gaussian-core
systems~\cite{Loui00a}.  A final extension is to consider explicitly
colloid-polymer mixtures, by determining the effective
hard-sphere/polymer potential along the lines set out in this paper.
The general methodology should, more generally, be applicable to
dilute and semi-dilute solutions of linear, branched or star polymers
in confined geometries.

\section*{Acknowledgements}
 AAL acknowledges support from the Isaac Newton Trust, Cambridge, PB
acknowledges support from the EPSRC under grant number GR$/$M88839,
EJM acknowledges support from the Royal Netherlands Academy of Arts
and Sciences.  We thank David Chandler, Daan Frenkel, Ludger Harnau,
Christos Likos, Hartmut L\"{o}wen, and Patrick Warren for helpful
discussions.

\appendix

\section{Depletion potential for ideal polymers}

In this Appendix we pursue a programme similar to that of the main
text, but now for the simpler case of ideal Gaussian polymer coils of
size $R_g$.  Consider two parallel walls of area $A = L_x L_y$ a
distance $L_z$ apart.  In the limit $L_x,L_y >> R_g$, the full
partition function for a single polymer is given by\cite{Asak54}:
\begin{equation}\label{eqA.1}
Z_1 = L_x L_y L_z \frac{8}{\pi^2} \sum_{p=1,3,...}^{\infty} \frac{1}{p^2}
\exp \left( -\frac{\pi^2 R_g^2 p^2}{L_z^2} \right).
\end{equation}
From this, various properties, such as the depletion interaction between
two walls, can be exactly calculated.
  
\begin{figure}
\begin{center}
\epsfig{figure=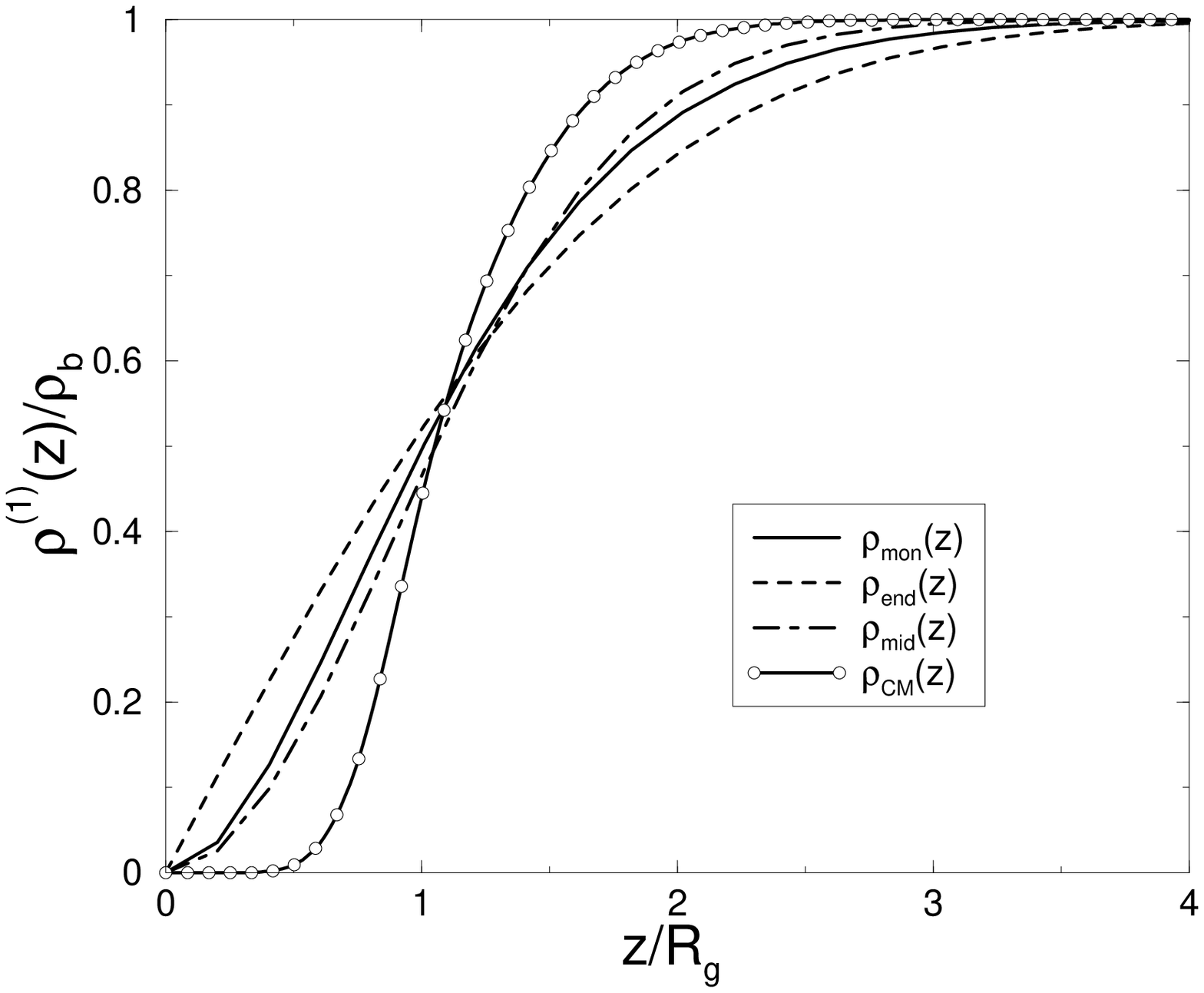,width=8cm} 
%\begin{minipage}{8cm}
\caption{\label{fig:end-mid} Normalized densities
$\rho^{(1)}(z)/\rho_b$ for ideal polymers near a wall. Also shown is
the normalized monomer density\protect\cite{Marq90} which falls, not
unexpectedly, between the end and mid-point densities.  The CM profile
is the steepest since the CM cannot easily approach the wall as
closely as the end or midpoints can. }
%\end{minipage}
\end{center}
\end{figure}

Similarly, from the underlying Green's function (see e.g. p. 19 of
ref.\cite{Doi86}), the polymer end-point and mid-point density
distributions near a single wall are found to be:
\begin{eqnarray}\label{eqA.2}
\frac{\rho^{(1)}_{end}(z)}{\rho_b} &=& \mbox{erf} \biggl( \frac{z}{2 R_g}
\biggl) \\ \label{eqA.2b}
\frac{\rho^{(1)}_{mid}(z)}{\rho_b} &=& \left(\mbox{erf} \biggl(
\frac{z}{\sqrt{2} R_g} \biggl)\right)^2.
\end{eqnarray}
As shown in Fig.~\ref{fig:end-mid}, both show a clear depletion layer
 where polymer configurations are entropically excluded near the wall.
 In principle the CM distribution could also be calculated, but we
 have not yet succeeded in finding an analytic expression for
 it. Instead a polymer lattice model simulation with $L=500$ was used
 to generate the $\rho^{(1)}_{CM}(z)$ depicted in
 Fig.~\ref{fig:end-mid}.

Using either the end-points, mid-points or the CM as the centers of
 ``effective particles'', an analogy with the full ``soft particle''
 picture can be made.  The particle-particle effective interaction is
 of course zero, while the effective particle-wall interaction can be
 inverted from the densities in Eqs.~\ref{eqA.2} and \ref{eqA.2b} with
 the result:
\begin{eqnarray}\label{eqA.3}
\phi^{(1)}_{end}(z) &=& -\ln \left[ \mbox{erf} \biggl( \frac{z}{2 R_g}
\biggl) \right]  \\ 
\phi^{(1)}_{mid}(z) &=& -2\ln \left[ \mbox{erf} \biggl( \frac{z}{\sqrt{2} R_g}
\biggl) \right].
\end{eqnarray}
The potential $\phi^{(1)}_{CM}(z)$ for the CM profile can be obtained
numerically.  In contrast to the soft-particle picture for interacting
polymers, the inversion here is trivial, since for ideal particles the
effective potential is simply the potential of mean force.  The
interaction $\phi^{(1)}_{CM}(z)$ for the Gaussian particles is in fact
quite similar to the potential of mean force for SAW polymers at 
infinite dilution, shown in Fig.~\ref{fig:ib}. But, whereas the 
wall-polymer interaction for interacting polymers changes with the 
bulk density, the $\phi^{(1)}(z)$ for Gaussian chains is independent
of density.

Within the effective particle picture, two ways of calculating the
interaction between two parallel walls are:

\noindent {\bf (1)} the {\bf Potential Overlap Approximation} (POA):
Here the partition function of the effective particles confined
between two walls, a distance $L_z$ apart, is calculated for a total
external potential given by:
\begin{equation}\label{eqA.4}
\phi(z) = \phi^{(1)}(z) + \phi^{(1)}(L_z-z)
\end{equation}
where $z$ is the distance from one of the walls.  For simple atomic or
molecular fluids this superposition approximation would be exact and
lead to the correct partition function and related equilibrium
properties.  However, for effective particles this is not necessarily
the case as we shall see later on.

\noindent {\bf (2)} the {\bf Density Overlap Approximation} (DOA): Here
the density between two parallel walls is approximated by the product
of the densities near a single wall:
\begin{equation}\label{eqA.5}
\rho_b \rho(z) = \rho^{(1)}(z) \rho^{(1)}(L_z -z) 
\end{equation}
In contrast to the POA, this approximation is incorrect even for
simple atomic or molecular systems, although it is sometimes a useful
first approximation.  On the other hand, for ideal particles the POA
and DOA approximations are equivalent.

The original Asakura Oosawa model\cite{Asak58} approximates the
density profile $\rho^{(1)}(z)$ next to a single wall by a step
function of range $R = R_g$.  The depletion potential is  then
calculated within the DOA.  This can be improved by adjusting the
range of the step-function such that it excludes exactly the same
amount of polymer as the true density profile. For flat walls this
implies a step-function of range $R = 2R_g/\sqrt{\pi}$.

\begin{figure}
\begin{center}
\epsfig{figure=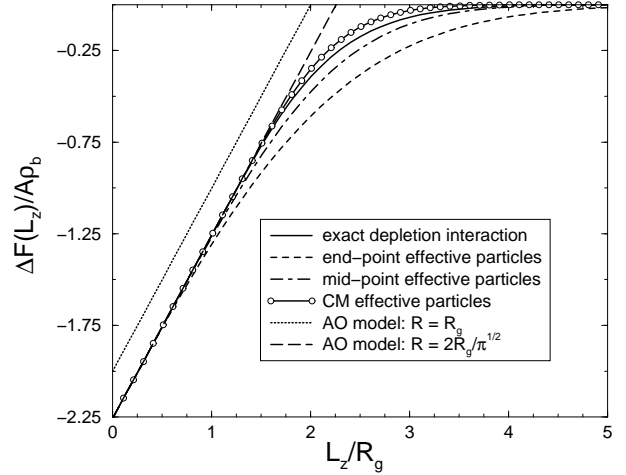,width=8cm} 
%\begin{minipage}{8cm}
\caption{\label{fig:vcompare} A comparison of the normalized depletion
interaction $\Delta F(L_z)/A\rho_b$ as a function of the wall-wall
separation $L_z/R_g$ for various approximations discussed in the text.  The
x-axis denotes.}
%\end{minipage}
\end{center}
\end{figure}

In Fig.~\ref{fig:vcompare} we plot the depletion  
free-energy induced by non-interacting polymers between two walls a
distance $L_z$ apart.  The exact expression was first calculated by
Asakura and Oosawa\cite{Asak54}; here we approximate it by the
following simple analytical expression: 
\begin{eqnarray}\label{eqA.6}
\frac{\Delta F (L_z)}{A} & = & \!
-\!\rho_b\! \left\{\!\!\frac{4}{\sqrt{\pi}}\!-\!L_z\!\!\left( 1 -
\frac{8}{\pi^2} e^{-\frac{\pi^2 R_g^2}{L_z^2}}\right)\!\! \right\}\!;\!L_z\!<\!4.332R_g \nonumber \\
\frac{\Delta F (L_z)}{A}  & = &  0;
\;\;\;\;\;\;\;\;\;\;\;\;   L_z > 4.332 R_g,
\end{eqnarray}
which arises from taking only the ``ground state'' of the partition
function in Eq.~(\ref{eqA.1}), and cutting the potential off where it
crosses $0$.  This approximation is so accurate that the difference
with the exact interaction cannot be resolved in
Fig.~\ref{fig:vcompare}.  The  effective particle representations,
based on end-points, mid-points or the CM, 
provide a fairly good approximation to the full depletion interaction,
while the two versions of the AO model do not perform as well.

\begin{figure}
\begin{center}
\epsfig{figure=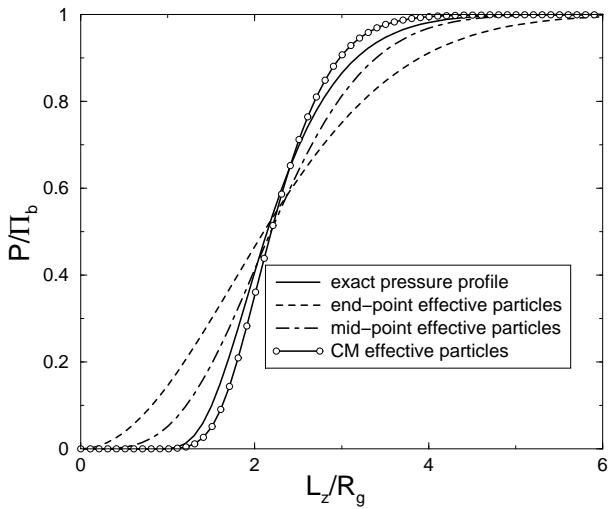,width=8cm} 
%\begin{minipage}{8cm}
\caption{ \label{fig:pcompare} Comparison of the reduced pressure
$P(L_z;\mu)/\Pi_b$  between the walls as a function of wall-wall
separation  $L_z/R_g$. $\Pi_b$ is the bulk pressure. }
%\end{minipage}
\end{center}
\end{figure}

The pressure profiles shown in Fig.~\ref{fig:pcompare} demonstrate
that the end and mid-point effective particle approaches underestimate
the steepness of the pressure profile, while their CM counterpart
overestimates the steepness.  The differences between the effective
particle representations and the exact results arise from our use of
the POA.  Because the underlying polymer configurations can easily
extend to distances greater than $2 R_g$, the POA, which implicitly
assumes that the interaction of an effective particle with one wall is
not directly affected by the presence of a second wall, begins to
break down for strong confinement.  Interestingly, the trend shown in
Fig.~\ref{fig:pcompare} for the CM effective particle representation
of the Gaussian coils mirrors the trend shown in
Fig.~\ref{fig:Pprofile} for interacting SAW polymers, suggesting that
the differences for the latter also arise from the breakdown of the
POA approximation.

We note that end-points or mid-points could also be used to construct
an effective particle picture of interacting polymer solutions.  For
example, the mid-point representation would be very similar to the
two-arm limit of a star-polymer, for which a number of results have
been recently derived\cite{vonF00,Liko00,Watz99}.  There are
differences with the CM representation; for example, the mid-point
equivalent of Eq.~(\ref{eq3.4}) would scale as\cite{Liko00,Witt86}:
\begin{equation}\label{eqA.7}
v_2(r) \sim -\frac{5 \sqrt{2}}{9} \ln\left(\frac{r}{R}\right)
\end{equation}
at short distances.  Here $R$ is a lengthscale proportional to the
polymer size $R_g$.  In contrast to the CM case, this interaction
diverges at full overlap so that one would expect some qualitative
differences in the behavior of the underlying soft particle fluids.
However, the two approaches should, in principle, produce similar
results for the thermodynamic properties of polymer solutions.  The
relative merits of using mid-points v.s.\ the CM to describe polymer
solutions are currently under investigation.

Finally we note that several results for ideal polymers, obtained here
by using end-point or mid-point densities, can be also obtained by
using the exact {\em monomer} density profiles near one
wall\cite{Marq90} together with the DOA approximation\cite{Tuin00}.

\end{multicols}


\begin{thebibliography}{99}
\bibitem{B2} See for example pp. 77,78 of ref.\protect\cite{deGe79},
where the hard-sphere analogy is attributed to Flory (note, however,
that this classic text was written before ref.\protect\cite{Gros82}).
In fact, $B_2$ scales with $R_g^3$ because that is the only
lengthscale, not because the polymers exclude each other in a
hard-sphere like fashion, a point which has also been emphasized by
Sch\"{a}fer\cite{Ohta82,Krug89}.
\bibitem{web}Parameters for the spline fits to the effective
potentials are available from the authors.
\end{thebibliography}
\end{document}